%% file: main.tex
\documentclass[reprint,twocolumn,amsmath,amssymb,aps,superscriptaddress]{revtex4-1}

\usepackage{editing}

\usepackage{graphicx}% Include figure files
\usepackage{dcolumn}% Align table columns on decimal point
\usepackage{bm}% bold math
\usepackage{color}
\usepackage{hyperref}% add hypertext capabilities
\hypersetup{colorlinks, citecolor=blue, linkcolor=blue, urlcolor=blue}

\usepackage{amsmath}
\usepackage{amssymb}

\usepackage{tikz}
\usepackage{standalone}

\usetikzlibrary{decorations.pathmorphing}
\usetikzlibrary{decorations.markings}
\usetikzlibrary{arrows,decorations}
\usetikzlibrary{shapes.misc, positioning}
\usetikzlibrary{shapes.arrows, fadings, calc}

\newcommand{\ket}[1] {| #1 \rangle}
\newcommand{\bra}[1] {\langle #1 |}

\newchange{}{DarkGreen}{OrangeRed}%
\changeOnOff

\begin{document}

%\usepackage{lineno}

%\preprint{APS/123-QED}

\title{Engineering Vibrationally Assisted Energy Transfer in a Trapped-Ion Quantum Simulator}

\author{Dylan J Gorman}
\affiliation{
Department of Physics, University of California, Berkeley, California 94720, USA
}
\author{Boerge Hemmerling}
\affiliation{
Department of Physics, University of California, Berkeley, California 94720, USA
}
\affiliation{
{\it Present address}: Department of Physics and Astronomy, University of California, Riverside, California 92521, USA
}
\author{Eli Megidish}
\affiliation{
Department of Physics, University of California, Berkeley, California 94720, USA
}
\author{Soenke A. Moeller}
\affiliation{
Department of Physics, University of California, Berkeley, California 94720, USA
}
\author{Philipp Schindler}
\affiliation{
Institut f\"ur Experimentalphysik, Universit\"at Innsbruck, Technikerstra\ss e 25, A-6020 Innsbruck, Austria
}
\author{Mohan Sarovar}
\affiliation{
Extreme-scale Data Science and Analytics,
Sandia National Laboratories, Livermore, California 94550, USA
}
\author{Hartmut Haeffner}
\affiliation{
Department of Physics, University of California, Berkeley, California 94720, USA
}

\begin{abstract}
Many important chemical and biochemical processes in the condensed phase are notoriously difficult to simulate numerically. Often this difficulty arises from the complexity of simulating dynamics resulting from coupling to structured, mesoscopic baths, for which no separation of time scales exists and statistical treatments fail. A prime example of such a process is vibrationally assisted charge or energy transfer. A quantum simulator, capable of implementing a realistic model of the system of interest, could provide insight into these processes in regimes where numerical treatments fail.
We take a first step towards modeling such transfer processes using an ion trap quantum simulator. By implementing a minimal model, we observe vibrationally assisted energy transport between the electronic states of a donor and an acceptor ion augmented by coupling the donor ion to its vibration.
We tune our simulator into several parameter regimes and, in particular, investigate the transfer dynamics in the nonperturbative regime often found in biochemical 
situations.

%We further study how our approach can be extended to system sizes where brute-force numerical methods become challenging.%, providing thus a useful tool to study transfer processes in complex systems.
\end{abstract}

\maketitle

\section{Introduction}

Charge and energy transfer are essential to many important processes in chemistry, biology, and emerging nanotechnologies. Such transfer processes often occur in noisy thermal environments that strongly modify the transfer dynamics and, in some cases, even improve the transport efficiency or robustness
\cite{Forster1959,Bruno1992,May2011,Nit-2006,Rebentrost2009,Caruso2009}.

A prominent example is the energy transfer from pigments in light-harvesting complexes towards reaction centers, where efficiency is believed to critically depend on the
spectral properties of the environment \cite{Adolphs2006,Chin2013,Irish2014,Kreisbeck2014,Nalbach2015,Dijkstra2015,Fujihashi2015,Liu2016}.
%Such noise- (or {\em environmentally}-) assisted transport processes are often difficult to study numerically owing to the complexity of their structured, mesoscopic molecular environments.

In these processes, the dominant sources of fluctuations and noise are often intramolecular vibrations and solvent dynamics \cite{Nit-2006}.
Understanding the influence of this molecular environment on transport dynamics requires solving complex and often fully quantized models that become intractable to theoretical treatments even for systems of moderate size. Moreover, direct experimental studies of these vibrationally assisted energy transfer (VAET) phenomena are difficult to interpret since the underlying mechanisms cannot be isolated.
A quantum simulation of such models, on the other hand, offers the possibility to both isolate and control the interesting aspects of the underlying mechanisms.

\begin{figure*}
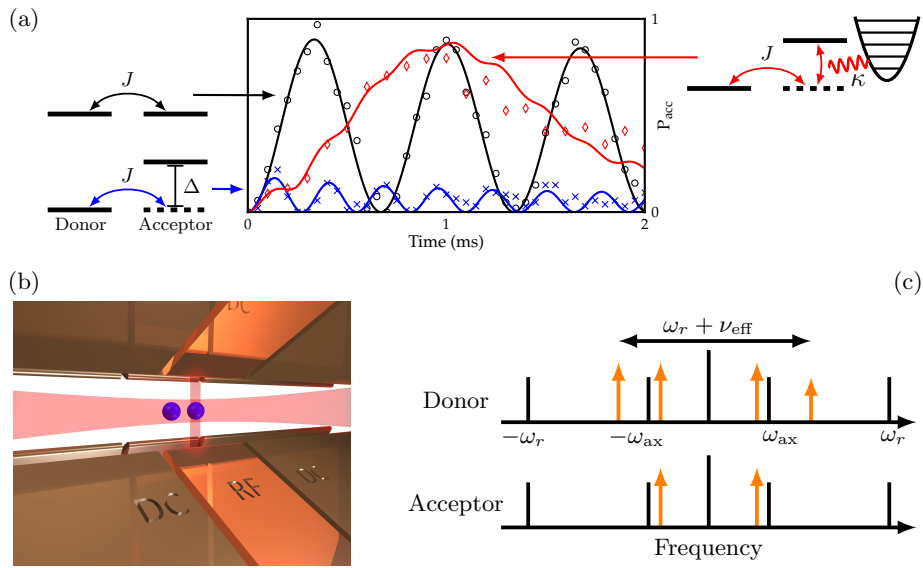


  \includestandalone{fig1box}
          %\includegraphics[width=1\columnwidth]{images/trap.png}
 %\internallinenumbers
  \caption{
  (a) Schematic illustrations of the VAET process
     and time dynamics of the target state population in various regimes. Without the presence of an environment ($\kappa = 0$, left drawings), the transition probability
    from donor to acceptor states
    is attenuated in the presence of an energy barrier $\Delta$. By coupling to an environment ($\kappa > 0$, right drawing), an excitation can move between donor and acceptor sites by exchanging
    energy with a phononic environment.
    Time traces illustrate the three situations:
     (1) $\Delta = 0$: black trace (theory) and data points ($\circ$).
    (2) $\Delta > J$, without assistance from the
    environment: blue trace (theory) and data points ($\color{blue}\times$).
    (3) $\Delta > J$
    with assistance from the environment (VAET process): red trace (theory) and data points ($\color{red}\diamond$).
    (b) Schematics of the ion trap and laser beams generating the simulated
    Hamiltonian. Internal levels of the ions (blue spheres) serve as energy sites.
    A laser beam illuminating both ions generates the site-site coupling with
    strength $J$.
    A localized beam generates the coupling to the environment with
    strength $\kappa$ and controls the detuning $\Delta$.
    (c) Laser tones (orange arrows) generating the simulated Hamiltonian. Vertical black lines represent
    available transitions, with the atomic resonance of the respective ion in the center. The detuning $\Delta$ is introduced via an ac-Stark shift by imbalancing the relative power of the two tones of the local beam, as indicated in the figure.
    \label{fig:energy-transfer}
    \label{fig:laserscheme}
    \label{fig:trap}
    }
\end{figure*}

In a simple model featuring vibrationally assisted energy transfer, the environment consists of a thermalized vibrational degree of freedom that can assist the exchange of quantized excitations between a donor and an acceptor site [see Fig.~\ref{fig:energy-transfer}(a)]. 
In general, these sites exhibit different energies such that transfer only occurs if the excess energy is taken up or provided by the vibration--as such, the environment assists in the transfer process. This model captures the important features of vibrationally enhanced phenomena, such as the dependence of transfer efficiency on the spectral properties and temperature of the environment.

%Instead of exploring these models numerically, one may be interested in studying them in quantum simulations \cite{Feynman1982,Lloyd1995}. However, so far analog quantum simulators have mainly been applied to studying closed quantum systems \cite{Blatt2012,Britton2012,Zhang2017,Bloch2012,Houck2012} and not open quantum systems as required in this application.

Here, we demonstrate VAET in isolation and under fully controlled conditions. 
We encode the VAET process in a trapped-ion quantum simulator, where energy transfer between the electronic states of
two ions is enhanced when coupled to an environment in the form of the
thermal vibrational motion of the ion crystal.
%We observe transfer processes wherein the environment changes by an integer number of motional quanta. 
We observe the hallmark feature of VAET, namely, the strong dependence of the energy transfer efficiency on the temperature and the spectral characteristics of the environment.
In addition, we tune our quantum simulator into nonperturbative parameter regimes, similarly to what is encountered in models of biochemical processes.

The model we consider couples two-level energy sites
[donor ($d$) and acceptor ($a$)]
with strength $J$ described by $\frac{J}{2} \sigma_x^{(d)} \sigma_x^{(a)}$ [see Fig.~\ref{fig:laserscheme}(a)].
In the absence of additional interactions,
this coupling will cause a single excitation to oscillate between the sites with frequency $J$. A relative energy detuning $\Delta$, represented by a term $\frac{\Delta}{2} \sigma_z^{(d)}$, imposes an energy cost to move an excitation between the sites, therefore inhibiting complete transfer of excitations. A long-lived vibrational mode in the environment is modeled as a harmonic oscillator with frequency $\nu_{\rm{eff}}$ and is coupled to the sites in the form of $\frac{\kappa}{2} \sigma_z(a + a^\dagger)$. Quantum mechanically, the role of the environment may be understood as providing an extra degree of freedom, which helps to satisfy energy conservation in the transfer process. For instance, if $\nu_{\rm{eff}} \approx \sqrt{\Delta^2+J^2}$, intersite
transfer can occur, provided the environment changes its vibrational quantum number by 1 [see Fig.~\ref{fig:laserscheme}(a)]. Higher-order processes also occur: If
$\nu_{\rm{eff}} \approx \sqrt{\Delta^2+J^2}/k$ for any integer $k$, intersite transfer occurs and the environment changes by $k$ vibrational quanta. 
A classical analogue to this process is that the intersite energy difference $\Delta$ is modulated, parametrically moving the excitation between the sites.
The resulting Hamiltonian for the VAET model is then
($\hbar = 1$)

\begin{equation}
  H =  \frac{J}{2} \sigma_x^{(d)} \sigma_x^{(a)}+ \frac{\Delta}{2} \sigma_z^{(d)} + \frac{\kappa}{2}\sigma_z^{(d)}(a + a^\dagger) + \nu_{\rm{eff}} a^\dagger a \; .
  \label{eqn:h-model}
\end{equation}

The remainder of this article is organized as follows. After discussing 
details of the implementation in Sec.~\ref{sec:implementation}, 
we present our experimental results in Sec.~\ref{sec:experiments}. Here, we first study the case where the detuning $\Delta$ is larger than the site-site coupling $J$ such that appreciable energy transfer can only occur with the assistance of the environment. Then, we make $\Delta$ comparable to $J$ and operate the simulator in a regime where all Hamiltonian terms compete with each other. 
%This is the situation most relevant to energy transfer in photosynthetic light harvesting complexes in a regime with no clear separation of the energy scales~\cite{Ish.Cal.etal-2010}. 
We then initialize the environmental mode in a thermal state with variable temperature; as such, this model corresponds to a biochemical scenario where the coupling of one pigment molecule to its environment is dominated by coupling to one slowly relaxing harmonic mode. 
We finish by providing an outlook and conclusions in Sec.~\ref{sec:outlook}.

\begin{figure*}
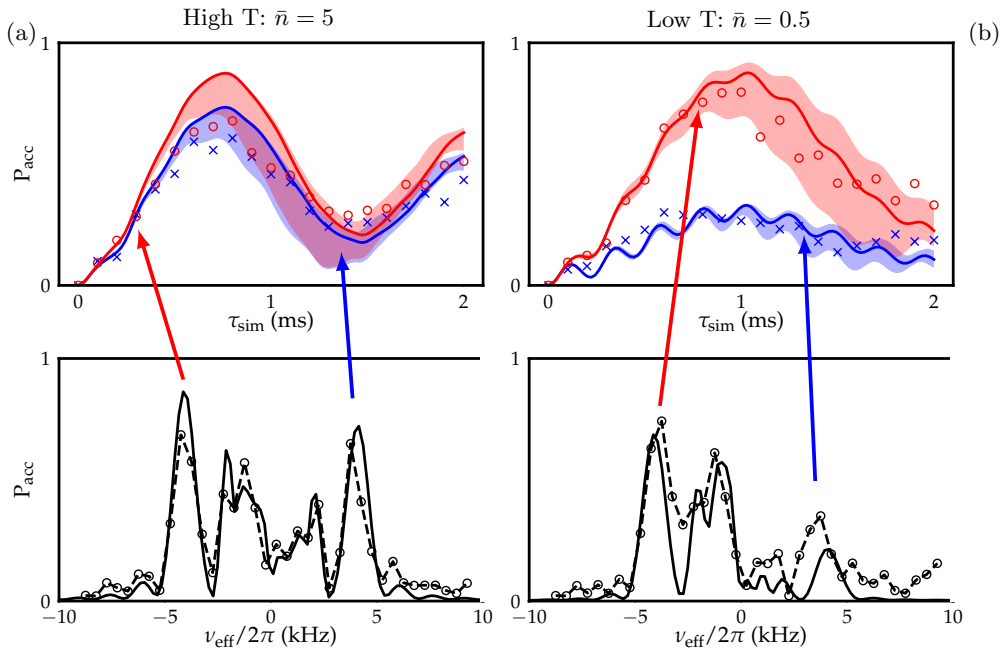

  \includestandalone{fig2box}
  \vspace*{-0.25cm}
  %\internallinenumbers
  \caption{
  Energy transfer probability to the acceptor $P_\textrm{acc}$
         %to the acceptor state 
         vs.~simulation time $\tau_{\rm{sim}}$ and vibrational
         frequency $\nu_{\rm{eff}}$.
         Left (a) [right (b)] plots show the time dynamics
         at environmental temperature $\bar{n} = 5$ ($\bar{n}=0.5$).
         Upper plots show the time dynamics $P_\textrm{acc}$ with $\nu_{\rm{eff}}/2\pi \approx +4.56$~kHz (blue points, $\color{blue}\times$)
   and $\nu_{\rm{eff}}/2\pi \approx -4.56$~kHz (red points, $\color{red}\circ$).
            The lower plots show $P_\textrm{acc}$ vs $\nu_{\rm{eff}}$, where the simulation time $\tau_{\rm{sim}}$ is fixed to $0.7$~ms.
         For all cases,
         $\{J, \kappa, \Delta\} = 2\pi \times \{1.30(1), 1.40(4), 4.56(2)\}$~kHz. Solid lines are numerical simulations of the system
         with all parameters determined through independent calibrations.
        The shaded regions represent the estimated systematic uncertainty on the theoretical curve, obtained from the measurement error in the calibration parameters.  Note that the theory estimates include the effect that the detuning $\Delta$ fluctuates around its mean due to relative intensity noise of 0.02 as well as due to variations of the ion-laser coupling strength via finite-temperature effects. In (a), the standard deviation of the corresponding distribution is  $2\pi \times 0.23$~kHz, whereas in (b) it is  $ 2\pi \times 0.1$~kHz. 
         %The deviation from theory in \textbf{(a)} can be explained by $|\nu_{\rm eff}-\Delta|\approx 330$~Hz, likely caused by a jump in the radial frequency  between taking the spectral scan and performing the time trace. Radial frequency jumps of $\sim300$~Hz are observed about once per hour in our %apparatus. 
         Solid lines in the spectral plots represent a
         numerical solution where a small frequency offset 
         adjusts for a systematic bias of the vibrational frequency measurements. 
         %\change{The systematic error is dominated by}{Deviations from the theory are mainly due to fluctuations in the AC Stark shift ($\Delta$) due to finite
         %temperature of the ions, as well as}{radial trap frequency
         %instabilities of about $2\pi\times 250$~Hz}
         %\change{ leading to the same uncertainty of the assisting mode's frequency.}{}
         In all plots, the statistical error is smaller than the markers.
         The measured data points in the spectral plots are connected with a dashed line to guide the eye. 
    \label{fig:time_delta}
    }
\end{figure*}

\section{Experimental implementation \label{sec:implementation}}
We implement this model in a trapped-ion
quantum simulator consisting
of two trapped $^{40}$Ca$^+$ ions confined in a radio-frequency Paul trap [Fig.~\ref{fig:trap}(b)]. Ca$^+$ has a ground $S_{1/2}$ orbital and a metastable $D_{5/2}$ orbital. The magnetic substates $\ket{S} (m_j=1/2)$ and $\ket{D} (m_j=1/2)$ form a qubit, addressed by an optical transition near 729~nm \cite{Haeffner2008}. The electronic states of the two ions play the role of the energy sites. The state $\ket{DS}$ ($\ket{SD}$) corresponds to a single excitation localized to the donor (acceptor) site.
The two-ion crystal has six normal vibrational modes, only two of which are relevant to implement the Hamiltonian in Eq.~(\ref{eqn:h-model}): the axial stretch mode with $\omega_{ax} \approx 2\pi \times 1.3$~MHz and the radial rocking mode at $\omega_r \approx 2\pi \times 2.1$~MHz.
The rocking mode serves as the thermally occupied bosonic environment in the simulation, while the axial stretch mode mediates the coupling between the donor and acceptor sites.
The additional vibrational modes of the ion crystal generate Hamiltonian
terms that rotate with frequencies of at least $2\pi \times 300$~kHz. 
We operate our simulator in the regime where all Hamiltonian parameters
are of order a few~kHz and therefore neglect these additional couplings.

The amplitude of each term in the model Hamiltonian [Eq.~(\ref{eqn:h-model})] is
controlled by adjusting the strength and frequency of various laser tones as
summarized in Fig.~\ref{fig:laserscheme}(c).
In particular, the site-site coupling $\frac{J}{2}\sigma_x^{(d)}\sigma_x^{(a)}$ is implemented by a M\o lmer-S\o rensen quantum interaction \cite{Sorensen1999} via
the axial vibrational mode. This interaction is generated by applying a global laser beam with tones detuned from the qubit transition by $\pm (\omega_{\rm ax} - \delta_{\rm ms})$, where $\delta_{\rm ms} = 2\pi \times
30 \rm{\; kHz} \ll \omega_{\rm ax}$ [see Fig.~\ref{fig:laserscheme}(c)].
The coupling strength is given by $J = \eta_{\rm ax}^2 \Omega_1^{\rm G} \Omega_2^{\rm G}/\delta_{\rm ms}$ \cite{Sorensen1999}, where $\eta_{\rm ax} \sim 0.05$ is the Lamb-Dicke parameter for this mode, and $\Omega_{1,2}^{\rm G}$ are the Rabi frequencies of the two laser tones. For calibrating the simulator, $J$ is measured independently by operating the simulator with  $\Delta, \kappa = 0$. We extract $J$ by fitting the population transfer as a function of $\tau_{\rm sim}$ to a sinusoid.
Note that in the single excitation manifold, spanned by $\ket{SD}$ and $\ket{DS}$, the $\sigma_x^{(d)}\sigma_x^{(a)}$ interaction is the same as the energy exchange interaction $\sigma_+^{(d)} \sigma_-^{(a)} + \sigma_-^{(d)} \sigma_+^{(a)}$, which is more commonly used in the context of charge and energy transfer in biochemical systems.

The site-environment coupling is engineered via a bichromatic laser beam localized to the donor ion.  The two tones of this beam are detuned from the optical transition by $-\omega_{\rm r}/2$ and $\omega_{\rm r}/2 + \nu_{\rm{eff}}$  \cite{Kim2008}. Note that $\nu_{\rm{eff}}$ is defined as the difference between the ion-crystal rocking-mode frequency $\omega_{\rm r}$ and the frequency splitting between the two laser tones. This generates the effective interaction $\frac{\kappa}{2} \sigma_z^{(d)}(a + a^\dagger) + \nu_{\rm{eff}} a^\dagger a$, where $\kappa = \eta_{\rm r} \Omega_1^{\rm L} \Omega_2^{\rm L}/\omega_r$. Here, $\Omega_i^{\rm L}$ is the on-resonance coupling between the $\ket{S}$ and $\ket{D}$ states generated by the $i$th tone of the local beam. The parameter $\eta_{\rm r} = 0.039(1)$ is the Lamb-Dicke parameter for the radial rocking mode. Experimentally, we adjust the laser powers of each of the tones to reach the desired coupling $\kappa$. 

To calibrate the coupling to the environment, $\kappa$, we measure the Rabi frequencies $\Omega_1^{\rm L}$ and $\Omega_2^{\rm L}$ directly in a two-stage procedure. First, both tones in the local beam are blue-shifted in a double-pass acousto-optic modulator (AOM) by  $\omega_{\rm r}/2$, such that one tone is resonant with the carrier transition. In that configuration, we measure the frequency of Rabi oscillations and extract $\Omega_1^{\rm L}$. Then, we shift both tones red by $\omega_{\rm r}/2$ and again measure Rabi oscillations,
extracting $\Omega_2^{\rm L}$.  

In order to maintain a stable environment frequency $\nu_{\rm eff}$, we actively stabilize the radial trap frequency to within $2\pi \times 250$~Hz over the duration of one time scan (several minutes), following a method detailed in Ref.~\cite{Johnson2016}. The rocking-mode frequency $\omega_{\rm r}$ is determined via optical spectroscopy on the $\ket{S}\rightarrow \ket{D}$ transition. However, ac-Stark shifts from the probe beam make a small correction on the order of a few hundred Hertz necessary. The systematic effect could be avoided by directly exciting the vibration with either a modulated optical radiation pressure force \cite{Ramm2014} or by measuring all center-of-mass modes using electrically oscillating fields. 

We control the detuning $\Delta$ by adjusting 
the power balance of the two tones on the local beam generating the site-bath coupling. The resulting detuning is measured by setting $\nu_{\rm{eff}} = 2\pi\times 30$~kHz, sufficiently far off-resonant such that the coupling to the vibrational mode can be neglected. In the case $\Delta \gtrsim J$,  the dynamics can be fit to the one given by the simplified  Hamiltonian $H = \frac{J}{2}\sigma_x^{(a)} \sigma_x^{(d)} + \frac{\Delta}{2}\sigma_z^{(d)}$, extracting $J$ and $\Delta$. When $\Delta \gg J$, the population transfer is too small to be fit for the detuning.  In this case, we measure $\Delta$ in a Ramsey-type experiment \cite{Chwalla2007}. First, a global $\pi/2$ pulse is applied to both ions. 
%Then, the local beam is enabled, generating a detuning $\Delta$. The global beam generating the site-site coupling (which does not generate differential Stark shifts) is left off so as
%not to interfere with the measurement.
After an interrogation time $\tau$, a second $\pi/2$ pulse is applied and the parity $P$ is recorded. The operator $P$ is defined as $P(\ket{SS}) = P(\ket{DD}) = 1$ and $P(\ket{SD}) = P(\ket{DS}) = -1$. The  parity $P(\tau)$ oscillates with frequency $\Delta$. In both cases, an analytical correction is applied to account for the small change in Stark shift (up to $2\pi\times 200$~Hz) arising from moving one laser tone by $2\pi\times 30$~kHz as necessary to decouple the bath from the dynamics.

The actual experimental procedure for the simulation is as follows: We start by Doppler cooling all the vibrational modes of the ion string to a mean occupation number of 6--12, followed by optical pumping both ions to the state $\ket{SS}$. We further cool the axial center-of-mass and stretch modes to the ground state via resolved sideband cooling \cite{Diedrich1989a}.  The assisting mode is then prepared via resolved sideband cooling \cite{Diedrich1989a} to an adjustable mean thermal occupation $0.04 \lesssim \bar{n} \lesssim 12$. The thermal occupation value is chosen by varying the duration of the cooling process. For small temperatures, the average population is extracted by comparing red- and blue-sideband excitation of the mode \cite{Turchette2000}, while for higher temperatures, the sideband strength is compared to the carrier transition strength. The donor is then excited via a local rotation, leading to the combined electronic state $\ket{DS}$. Then, the local and global laser beams generating the model Hamiltonian are applied for a time $\tau_{\rm{sim}}$. Finally, the combined electronic state of both ions is measured by recording the fluorescence with a charged-coupled device (CCD) camera \cite{Dehmelt1975}.
%by an electron shelving measurement \cite{Dehmelt1975}.
For each parameter setting $\{J$, $\kappa$, $\Delta$, $\nu_{\rm eff}\}$, the simulation is run 100--500 times. The transfer probability is then given by the population in the state $\ket{SD}$.
For all data, we report the conditional probability $P_\textrm{acc}$ that the system has undergone state transfer, i.e.~$P_\textrm{acc} = P_{SD}/(P_{SD} + P_{DS})$. This corrects for an average population loss
from the simulation subspace of 10\%
arising from imperfect state preparation and incoherent excitation of optical
transitions due to spectral impurities in our qubit laser at 729~nm.

In addition to probing the time dynamics of the VAET process by varying the simulation time $\tau_{\rm{sim}}$, we also investigate its spectroscopic properties. For the spectroscopic measurements,  we measure the energy transfer probability for fixed $\{J$, $\kappa$, $\Delta\}$ and simulation time $\tau_{\rm sim}$, varying the frequency $\nu_{\rm{eff}}$.

\begin{figure*}
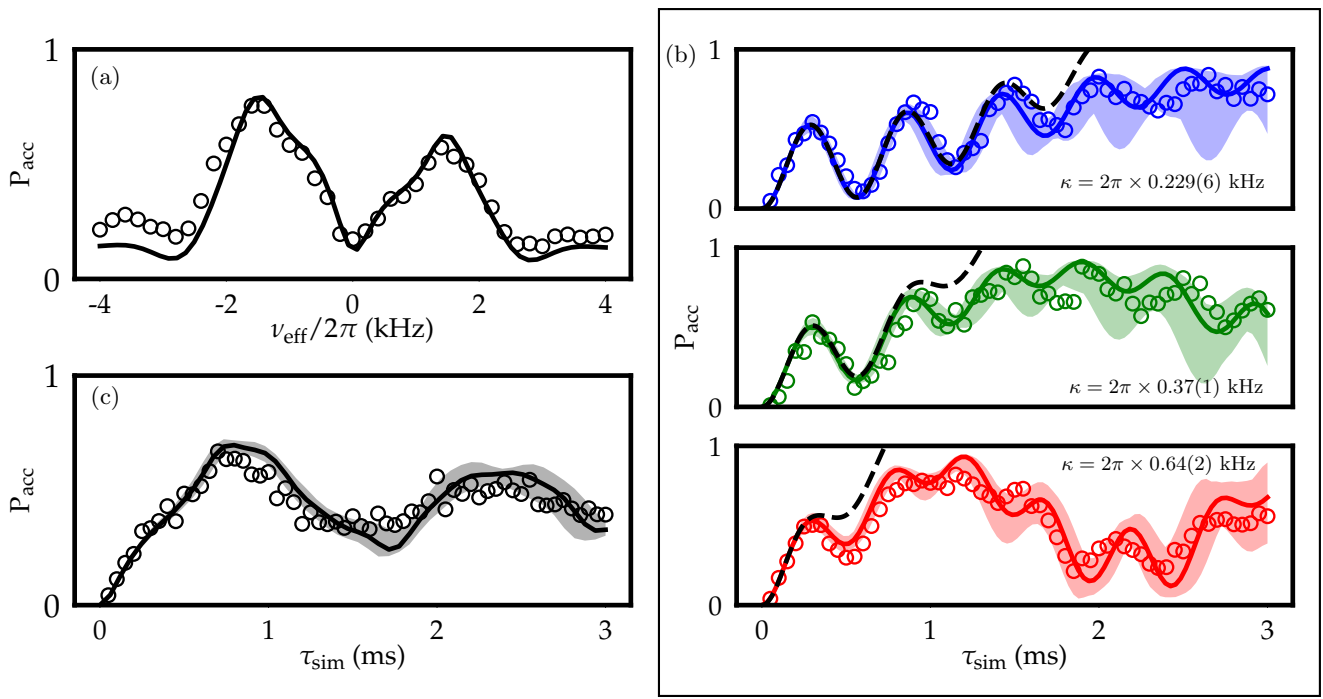

   \includestandalone{fig3box} % include standalone throws and error
   %\includegraphics{fig3box}
   %\internallinenumbers
   \caption{   
   (a) Energy transfer
	  probability $P_\textrm{acc}$
   vs.~vibrational frequency $\nu_{\rm{eff}}$
      in the small detuning regime with
      $\{J, \kappa, \Delta\} = 2\pi \times \{1.22(3), 0.63(2), 1.226(3)\}$~kHz, $\tau_{\rm{sim}} = 0.7$~ms and $\bar{n} = 2.7$. 
      Direct coupling between the sites exceeds the energy differences between the various multiphonon processes and only two peaks
      remain, corresponding to either subtracting or adding energy.
          (b) $P_\textrm{acc}$ vs~simulation time for different environmental couplings $\kappa$
          for low environment temperature ($\bar{n} = 0.04$).
          In all three scans, all parameters except $\kappa$ are constant, $\{J, \Delta, \nu_{\rm{eff}}\} = 2\pi \times \{1.27(5), 1.27(8), -1.72(4)\}$~kHz. Coupling
          $\kappa$ increases from top to bottom as
          indicated.
          %$2 \pi \times$ (0.229(6), 0.37(1), 0.64(2))~kHz.
         Solid traces are numerically simulated dynamics with all parameters determined by independent measurements. Dashed black traces correspond to the predictions of the perturbative treatment developed in Ref.~\cite{VAETSupplemental}. 
      (c) $P_\textrm{acc}$ vs~simulation time $\tau_{\rm{sim}}$
      at high temperature ($\bar{n} = 12$ quanta). Simulation parameters:
       $\{J, \kappa, \Delta, \nu_{\rm{eff}}\} = 2\pi \times \{1.17(3), 0.63(2), 1.59(3), -1.72(6)\}$~kHz. In the time traces, the shaded regions represent the estimated systematic error on the theoretical curve, obtained from the measurement error in the simulation parameters. In all plots, the statistical error is smaller than the markers.
    \label{fig:temperature-time}
    \label{fig:k_groundstate}
    \label{fig:nu_scan}
       }
  
\end{figure*}

\section{Energy transfer dynamics}\label{sec:experiments}

VAET is especially well illustrated in the regime where the detuning $\Delta$ is larger than the coupling $J$. Under these conditions, the energy transfer
is suppressed without the assistance of the environment [Fig.~\ref{fig:time_delta}(a)]. Significant energy transfer occurs only at appropriate environmental frequencies corresponding to processes $\ket{SD, n} \rightarrow \ket{DS, n \pm k}, k \geq 1$, where $k$ vibrational quanta are either removed from or absorbed by the environment.
The most pronounced energy transfer is observed
at $\nu_{\rm{eff}} = \pm 2\pi \times 4$~kHz, where the occupation of
the environment changes by 1.
The unresolved peaks at smaller vibrational frequencies correspond to
multiphonon processes where the energy gap is bridged by several vibrational quanta.
The physical interpretation of $\nu_{\rm{eff}}<0$ is a sign change of the detuning $\Delta$ and corresponds to the situation where the environment absorbs the excess energy.

With the environment prepared near
the ground state, the vibrations
can take up excess energy, but they cannot
provide quanta in the transfer process.
As a result, the spectral scans in Fig.~\ref{fig:time_delta}(b),
where the environment contains an average of 0.5 quanta,
show a suppression for positive $\nu_{\rm{eff}}$ but exhibit strong peaks at negative $\nu_{\rm{eff}}$.
This asymmetry is expected only when the environment is prepared close to the ground state and together with the quantized nature of the signal, it is a signature of the quantum-mechanical nature of the environment.

To study these processes in more detail,
we measure the time dynamics of the
single-phonon resonances ($\nu_{\rm{eff}} = \pm 2\pi\times 4$~kHz). In the low-temperature
setting, the time for maximum energy transfer
to the acceptor state is around 1~ms,
corresponding to a characteristic time scale
of $\frac{J \kappa}{2 \Delta}$
given by a perturbative treatment (see Ref.~\cite{VAETSupplemental}). The dynamics is faster at higher temperatures
owing to stronger fluctuations from the environment. The previously mentioned asymmetry between the
positive and negative frequency peaks is also present in the low-temperature
time dynamics.

For sufficiently large detuning $\Delta$, the multiphonon processes $\ket{SD, n} \rightarrow \ket{DS, n + k}$
can each be spectrally resolved up to a maximum integer
value.
However, when $\Delta$ becomes comparable to $J$, these processes cannot be distinguished, as shown in Fig.~\ref{fig:nu_scan}(a).
In this regime, the energy sites are partially hybridized because of the intersite coupling, and significant energy transfer
occurs even without a coupling to the environment.
Here, direct intersite coupling, single-phonon, and multiphonon processes all contribute simultaneously to the transfer dynamics. 
This parameter regime, where $J\sim \kappa \sim \Delta$, is most relevant to energy transfer dynamics in photosynthetic light-harvesting complexes ~\cite{Ish.Cal.etal-2010}. We note that, exactly in this regime, approximate methods, which typically proceed by perturbation in one of these parameters, are not effective (see Ref.~\cite{VAETSupplemental}).

When the environment is cooled near the ground state, we observe almost complete energy transfer between donor and acceptor states [see Fig.~\ref{fig:k_groundstate}(b)]. This energy transfer is accelerated with increasing site-environment coupling $\kappa$. Similarly, we see that the perturbative treatment developed in Ref.~\cite{VAETSupplemental} breaks down at earlier times. We note that the non-Markovianity of the environment is particularly evident for the case of $\kappa=2\pi\times 0.64$~kHz, since the energy returns to the acceptor during the measurement time. Finally, in Fig.~\ref{fig:temperature-time}(c), we show time dynamics at higher temperatures [$\bar{n}=12$, with all other parameters similar to the lowest time trace in Fig.~\ref{fig:k_groundstate}(b)]. The overall trend in population transfer is preserved; however, the coherent oscillations are damped at the higher temperature.

\section{Conclusions and outlook \label{sec:outlook}}

Realistic models of chemical and biological environments require extending the simple model above to incorporate larger numbers of sites and environmental modes. 
However, as the number of sites and vibrational modes that must be accounted for increases, these models quickly challenge analytical and numerical methods. This is particularly the case in regimes where a separation of time scales does not exist and perturbative approximations are invalid. Moreover, it may be necessary to include higher excitations of the environment. In particular, vibrationally assisted processes are significant when the vibrational modes are almost resonant with the electronic energy differences, typically of order 100--200~cm$^{-1}$ in photosynthetic systems.
At room temperature, the relevant vibrations are each excited with mean phonon numbers on the order of one to two quanta.
Capturing the associated Boltzmann distribution via direct numerical simulation
may require truncating each harmonic oscillator Hilbert space above five quanta -- the computational resource equivalent of 2--3 qubits.
Thus, an $N$-site model with two vibrational modes per site to capture at least some aspects of the local spectral density would require a Hilbert space size comparable to  $(1+2\times2)N = 5N$ qubits. To model
the local environment more accurately, a Hilbert space with substantially more dimensions would be required. Currently, it is impossible to study the general dynamics in a Hilbert space equivalent to 50~qubits~\cite{Haener2017}. Extending our experimental
platform, it may be possible to encode the 10-site dynamics of our model in a
10-ion crystal, thereby outperforming brute-force classical computation. More sophisticated numerical techniques for simulating vibrational wave-packet dynamics have been developed recently, \emph{e.g.}, the multilayer multiconfiguration time-dependent Hartree method (ML-MCTDH) \cite{Schulze2016}. While such methods substantially reduce the computational burden of computing nonequilibrium dynamics substantially, their computational cost is still exponential with the number of vibrational modes tracked; hence, they eventually scale badly.

Even without outperforming classical resources, our platform can be extended
to study more qualitatively interesting physics. For example, by applying
phase modulation to the site-environment coupling beams, a broadband bath can
be implemented in the simulation. Damping can be implemented by adding sideband cooling of particular vibrational modes using auxiliary ions. In addition, moving to a three-site model
would allow the investigation of phenomena such as quantum ratcheting~\cite{Myers2015}. Finally, we note that one could also study steady-state dynamics of VAET by continuously exciting some of the ions on the $\ket{S_{1/2}}\leftrightarrow\ket{D_{5/2}}$ transition while providing a sink to other ions using light on the   $\ket{D_{5/2}}\leftrightarrow\ket{P_{3/2}}$ transition.
Thus, it may be possible to simulate realistic models of energy transfer processes in light harvesting processes or similar transport phenomena such as present in
organic electronic devices including solar cells \cite{Bakulin2012,Gelinas2013,vanHal2001,Smith2015} or as discussed in the context of olfaction \cite{Solovyov2012} and neuroreceptor activation \cite{Hoehn2015}.

In conclusion, we have implemented an analog quantum simulation of vibrationally assisted energy transfer using trapped ions. We further demonstrated tuning of the simulator from perturbative to nonperturbative regimes (see Ref.~\cite{VAETSupplemental} for a more detailed discussion). The latter case is particularly interesting, as VAET dynamics with larger structured environments in this regime becomes inaccessible to numerical treatment on current high-performance computers. We expect that our platform will be capable of simulating complex models, including larger structured environments, with various experimental advances.

Recently, we became aware of related work carried out at ETH Z\"urich \cite{Potocnik2017}.

\begin{acknowledgments}
This work has been supported by AFOSR through Grant No.~FA9550-15-1-0249 and by the NSF grant \#PHY 1507160. We thank Clarice Aiello, Jonathan Ouellet, and Birgitta Whaley for insightful discussions.
Sandia National Laboratories is a multimission laboratory managed and operated by National Technology and Engineering Solutions of Sandia, LLC, a wholly owned subsidiary of Honeywell International, Inc., for the U.S. Department of Energy's National Nuclear Security Administration under Contract No.~DE-NA-0003525.
% H.H., D.J G., P.S.~and M.S.~had the idea for the experiment.
% D.J G., B.H.~and E.M.~and S.A.M.~carried out the measurements.
% D.J G., B.H.~and E.M.~analysed the data and performed exact numerically simulation.
% All authors wrote the main part of the manuscript.
% M.H.~wrote the Supplemental Information.
% All authors contributed to the discussions of the results and manuscript.
% The authors declare that they have no
% competing financial interests.
% Correspondence and requests for materials
% should be addressed to hhaeffner@berkeley.edu.
\end{acknowledgments}

\bibliographystyle{apsrev4-1}
\bibliography{mendeley}

\newpage

%%%%%%%%%%%%%%%%%%%%%%%%%%%%%%%%%%%%%%%%%%%%%%%
% everything from here goes to the supplement %
%%%%%%%%%%%%%%%%%%%%%%%%%%%%%%%%%%%%%%%%%%%%%%%

\section{Supplemental Information}

\subsection{Perturbative and non-perturbative regimes of VAET}

We have experimentally simulated the minimal model exhibiting VAET, given by Eq.~1 of the main text, in a variety of parameter regimes. These include regimes where all terms of the Hamiltonian are of similar magnitude, hindering approximate treatments of the dynamics. While brute-force numerical simulations of the model in this regime are possible, these calculations quickly become intractable with increasing complexity of the model. In this section, we study perturbative regimes of the VAET model with a single long-lived vibrational mode to examine when the approximations are no longer applicable and an exact simulation is necessary. 

The exact dynamics of the system is given by
\begin{equation}
%\begin{align}
	\varrho(t) = U(t) \varrho(0) U^\dagger(t),
	\label{eq:Ut}
%\end{align}
\end{equation}
where $\varrho(t)$ is the density matrix describing the system, $\varrho(0) = \ket{DS}\bra{DS}\otimes \omega(\bar{n})$ is the initial state and $\omega(\bar{n})$ is the thermal state of a bosonic mode with average occupation $\bar{n}$. The time propagation operator is given by 
%\begin{equation}
\begin{align}
	U(t) &= \exp\left( -i\left[\frac{\Delta}{2} \sigma_z^{(d)} + \frac{J}{2}\sigma_x^{(d)}\sigma_x^{(a)} + \right.\right.\nonumber \\
	&~~~~~~~~~~~~~~~~~\left.\left.\frac{\kappa}{2}\sigma_z^{(d)}(a+a^\dagger) + \nu_{\rm eff}a^\dagger a\right] t\right).
\end{align}
%\end{equation}
Since the dynamics are confined to the $\{\ket{DS},\ket{SD}\}$ manifold, the Hamiltonian can be simplified as
\begin{equation}
%\begin{align}
	H_r = \frac{\Delta}{2}\sigma_z + \frac{J}{2} \sigma_x + \frac{\kappa}{2}\sigma_z(a + a^\dagger) + \nu_{\rm eff}a^\dagger a\,,
%\end{align}
\end{equation}
where $\sigma_\alpha$ are the Pauli matrices. The dynamics are then described by Eq.~(\ref{eq:Ut}), where $\varrho$ is a density matrix representing the state in the relevant two-dimensional manifold and the vibrational mode, with initial state $\varrho(0) = \ket{0}\bra{0}\otimes \omega(\bar{n})$, and $U(t) = \exp\left(-iH_r t\right)$.

We transform into an interaction picture with respect to all terms in the Hamiltonian except the site-environment coupling term; \textit{i.e.}, $\tilde{A}(t) = U_I(t)A U^\dagger_I(t)$ for any operator $A$, with the tilde denoting an operator in the interaction picture, and $U_I(t) = \exp\left( i\left[\frac{\Delta}{2}\sigma_z + \frac{J}{2} \sigma_x + \nu_{\rm eff}a^\dagger a\right]t\right)$. The coupling Hamiltonian then takes the form
\begin{equation}
%\begin{align}
	\tilde{V}(t) = \frac{\kappa}{2}\tilde{\sigma}_z(t)(ae^{-i\nu_{\rm eff}t} + a^\dagger e^{i\nu_{\rm eff}t})\,.
	\label{eq:int_V}
%\end{align}
\end{equation}
The propagator generating time dynamics in this frame is then
\begin{equation}
%\begin{align}
	U_{\rm int}(t) = \mathcal{T}\exp\left( -i\int_0^t ds \tilde{V}(s)\right),
%\end{align}
\end{equation}
where $\mathcal{T}$ denotes time-ordering.
The final quantity of interest is the population in the $\ket{SD}$ state, which is the $\ket{1}$ state in the simplified two-dimensional manifold, \textit{i.e.}
\begin{equation}
%\begin{align}
	P_1(t) = P_\textrm{acc} = \textrm{tr}\left(\tilde{\Pi}_1(t) U_{\rm int}(t) \varrho(0) U_{\rm int}^\dagger(t)\right)\,,
%\end{align}
\end{equation}
where $\tilde{\Pi}_1(t)$ is the projector $\ket{1}\bra{1}$ in the interaction picture.

\subsection{Perturbation in $\kappa$}
The most common approximation in open quantum system models is that the system-environment interaction $\kappa$ is small. In this regime, master equations, perturbative in $\kappa$, are often used to describe the system dynamics \cite{Bre.Pet-2002,Nit-2006}. However, we cannot resort to master equations in this case because the vibrational mode is not actively damped in the VAET model described by Eq.~1 in the main text. The absence of damping renders the Markovian approximation of a memoryless environment, and therefore master equation approaches, invalid.

\begin{figure*}
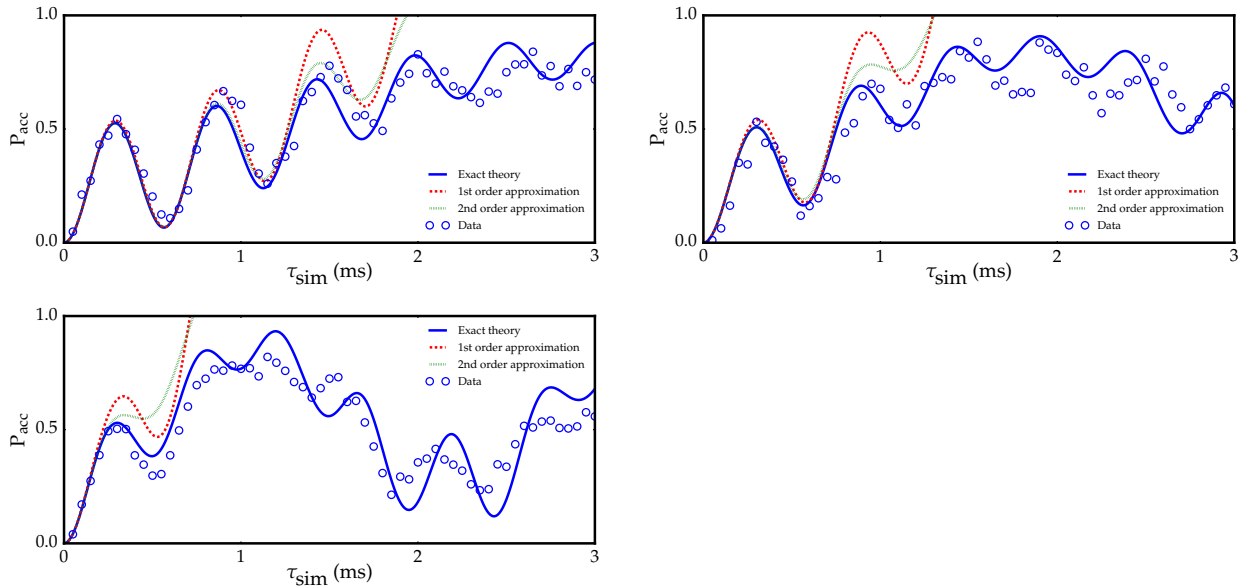

  \includestandalone{supplfig}
  %\internallinenumbers
  \caption{
  \label{fig:pert_fig} A comparison of the data shown in Fig.~\ref{fig:temperature-time}(b) of the main text, the exact solution of the VAET model, and first- and second-order approximations in $\kappa$ to the dynamics.}
\end{figure*}

An approximation of the dynamics can still be obtained in the small $\kappa$ regime by performing a perturbation expansion of the interaction picture propagator
\begin{align}
%\begin{align}
	U_{\rm int}(t) &= \mathcal{T}\exp\left( -i\int_0^t ds \tilde{V}(s)\right) \nonumber \\ &=\sum_{n=0}^\infty (-i)^n \int_0^t dt_1 ... \int_0^{t_{n-1}} dt_n \tilde{V}(t_1)\tilde{V}(t_2)...\tilde{V}(t_n)\,.\nonumber
\end{align}
%\end{align}
While we can calculate this quantity to any order in the perturbation expansion, the increasing computational expense renders this approach impractical. Fig.~\ref{fig:pert_fig} shows various perturbative approximations to the dynamics shown in Fig.~3(c) of the main text. As can be seen from this figure, this perturbative approximation always breaks down at some timescale, and this breakdown happens earlier as $\kappa$ increases. Our experimental simulations extend into a regime not described by this perturbative treatment. A more sophisticated perturbation expansion in the system-environment interaction is possible by first performing a polaron transformation of the VAET Hamiltonian (the effective system-environment coupling is reduced in the polaron frame) \cite{LeeMoixCao2012}. However, such expansions will also fail to desribe sufficiently long-time dynamics.

\subsection{Large $\Delta$ regime}
Another regime where one can develop a simpler picture of the VAET dynamics is when $\Delta$ dominates over the other parameters ($J,\kappa$).
To see this, we write out the explicit form of the the interaction picture representation of $\sigma_z$: $\tilde{\sigma_z}(t) = f_x(t)\sigma_x + f_y(t)\sigma_y + f_z(t)\sigma_z$, with
\begin{eqnarray}
%\begin{align}
    f_x(t) &=& \frac{\Delta J}{\Delta^2+J^2}\left(2 - e^{-i\Omega t} - e^{i\Omega t}\right) \nonumber \\
    &=& \frac{(J/\Delta)}{1+\left(J/\Delta\right)^2}\left(2 - e^{-i\Omega t} - e^{i\Omega t}\right)\nonumber \\
    f_y(t) &=& \frac{i J}{\sqrt{\Delta^2+J^2}}\left(e^{-i\Omega t} - e^{i\Omega t}\right) \nonumber \\
    &=& \frac{i (J/\Delta)}{\sqrt{1+\left(J/\Delta\right)^2}}\left(e^{-i\Omega t} - e^{i\Omega t}\right) \nonumber \\
    f_z(t) & =& \frac{1}{\Delta^2+J^2}\left[2\Delta^2 + J^2\left(e^{-i\Omega t} + e^{i\Omega t}\right)\right] \nonumber \\
    &=& \frac{1}{1+\left(J/\Delta\right)^2}\left[2 + \left(J/\Delta\right)^2\left(e^{-i\Omega t} + e^{i\Omega t}\right)\right]\nonumber,
%\end{align}
\end{eqnarray}
where $\Omega \equiv \sqrt{\Delta^2+J^2}$ and in the second equality of each term we have simply rewritten the expression in terms of $(J/\Delta)$.

Then the propagator in the interaction frame can be explicitly written as
%\begin{equation}
\begin{align}
    U_{\rm int}(t) =& \exp\Big( -i(F^-_x(t)\sigma_x + F^-_y(t)\sigma_y + F^-_z(t) \sigma_z)a \nonumber \\
    &~~~~ -i(F^+_x(t)\sigma_x + F^+_y(t)\sigma_y + F^+_z(t) \sigma_z)a^\dagger \Big),
    \label{eq:prop_explicit}
\end{align}
%\end{equation}
where 
\begin{equation}
%\begin{align}
    F^{\pm}_\alpha(t) = \frac{\kappa}{2}\int_0^t ds f_\alpha(s) e^{\pm i \nu_{\rm eff} s}, ~~~~~~ \alpha =x,y,z\,.
%\end{align}
\end{equation}
The key observation now is that amplitude of vibration-assisted transitions from donor to acceptor will be proportional to powers of $F^{\pm}_{x}$ and $F^{\pm}_{y}$. We will now examine one of these terms, $F^{-}_x(t)$, but the following arguments apply to all of these terms that quantify the magnitude of vibration-assisted transitions. Expanding $F^{-}_x(t)$,
%\begin{equation}
\begin{align}
    F^{-}_x(t) =& \frac{\kappa (J/\Delta)}{2(1+(J/\Delta)^2)} \times \nonumber \\
    &\int_0^t ds (2e^{-i\nu_{\rm eff} s} - e^{-i(\Omega+\nu_{\rm eff})s} - e^{i(\Omega-\nu_{\rm eff})s})\,.
    \label{eq:f-x}
%\end{equation}
\end{align}

There have been no approximations till this point. Now consider the case where $\Delta$ dominates over $J$ and $\kappa$ (\textit{e.g.}, Fig.2 in the main text). Then the prefactor $\frac{\kappa (J/\Delta)}{2(1+(J/\Delta)^2)} \approx \frac{\kappa (J/\Delta)}{2}$ is small and due to the oscillatory nature of the integral, $|F^-_x(t)|$ is small for $t\gg 1/\Omega, 1/\nu_{\rm eff}$, except for when $\nu_{\rm eff} \approx \pm\Omega$. This explains the resolved peaks in Fig. 2 of the main text; \textit{i.e.}, for large $\Delta$, vibration-assisted transitions are only significant when $\nu_{\rm eff} \approx \pm n\Omega$   since all powers of $F^-_x(t)$ (and the other coefficients) are generated by the exponential in Eq.~\ref{eq:prop_explicit}.

In this regime, due to this strong resonance condition, one can develop simple few-level models that approximate VAET dynamics.

\subsection{The non-perturbative regime of VAET}
Finally, consider the deeply non-perturbative regime of the VAET model where $J \sim \Delta \sim \kappa$. This is the situation
depicted in most of the panels in Fig.~3(a) and 3(b) of the main text. In this case, the perturbative expansion of the propagator in $\kappa$ is not valid and the approximations based on
small $J/\Delta$ do not hold. Consequently, the prefactor $\frac{\kappa (J/\Delta)}{2(1+(J/\Delta)^2)}$ of the expression for $F^-_x(t)$ in Eq.~\ref{eq:f-x} is not small. The
oscillating integrals still decay for values of $\nu_{\rm eff}$ far away from $\pm \Omega$, but the larger prefactor means that the vibrational frequency must be further away from resonance before
$|F^-_x(t)|$ becomes negligible. This explains the broader peaks in the spectrum shown in Fig. 3(a) of the main text.

For these parameters, we must simulate the full model for accurate predictions at timescales longer than $1/\Omega$ and $1/\nu_{\rm eff}$.

We note that this regime, where the resonance condition is violated, resembles the breakdown of the rotating-wave-approximation in the Rabi model (which makes the simplification to the Jaynes-Cummings model invalid) \cite{Wang2015a}.

\end{document}

%% file: fig1box.tex
\begin{tikzpicture}[
	scale = 1.0,
    carrier/.style={>=latex,draw=black,fill=black},
    rsb/.style={>=latex,draw=black,fill=black},
    bsb/.style={>=latex,draw=black,fill=black},
    %sigmaz/.style={->, >=latex,draw=orange,fill=orange, dash pattern=on 3pt off 1pt},
    %ms/.style={->, >=latex,draw=orange,fill=orange, dash pattern=on 2pt off 1pt},
	sigmaz/.style={->, >=latex,draw=orange,fill=orange},
    ms/.style={->, >=latex,draw=orange,fill=orange},
]

\def\wspaceA{1}
\def\wspaceB{3}
\def\carrierheight{1.2}
\def\sbheight{0.75}
\def\laserheight{1.0}

\def\mylw{0.5mm}

\begin{scope}[transform canvas = {scale = 0.85}, xshift = -2.5cm, yshift = 0.0cm]
   %\begin{scope}

   \begin{scope}[xshift = -9.25cm-3cm, yshift = 4.5cm]

	  \def\mylevelw{0.75mm}

	  %%%%%%%%%%%%%%%%%%%%%%%%%%%
	  % resonant scheme
	  %%%%%%%%%%%%%%%%%%%%%%%%%%%

	  \begin{scope}[xshift = 5cm, yshift = 0.5cm]
		 \draw[thick, line width = \mylevelw] (1.5,0) -- ++(1,0) node[midway, above=-2.5mm] (levelA) {};
		 \draw[thick, line width = \mylevelw, color = black] (0,0) -- ++(1,0) node[midway, above=-2.5mm] (levelB) {};
		 \draw[<->, >=latex, thick, color = black] (levelA) edge[out=135,in=45] node [above = 0.25mm, color = black] {$J$} (levelB);
		 %\draw[thick, line width = \mylevelw, color = black] (0,0.75) -- ++(1,0)  node[midway, below=-2.5mm] (levelC) {};
		 %\node at (3, 0.5) {$\kappa = 0$};

		 %   \node at ($(levelB.north)+(0.0, 1.1-0.25)$) {{\footnotesize donor\phantom{p}}};
		 %   \node at ($(levelB.north)+(2.0-0.5, 1.1-0.25)$) {{\footnotesize acceptor}};

	  \end{scope}

	  %%%%%%%%%%%%%%%%%%%%%%%%%%%
	  % delta scheme
	  %%%%%%%%%%%%%%%%%%%%%%%%%%%

	  \begin{scope}[xshift = 5.cm, yshift = -1cm]

		 \draw[thick, line width = \mylevelw] (0.0,0) -- ++(1,0) node[midway, above=-2.5mm] (levelA) {};
		 \draw[thick, line width = \mylevelw, dash pattern=on 3pt off 3pt, color = black] (1.5,0) -- ++(1,0) node[midway, above=-2.5mm] (levelB) {};
		 \draw[<->, >=latex, thick, color = blue] (levelA) edge[out=45,in=135] node [above = 0.25mm, color = black] {$J$} (levelB);
		 \draw[thick, line width = \mylevelw, color = black] (1.5,0.75) -- ++(1,0)  node[midway, below=-2.5mm] (levelC) {};
		 \node at (-3.5, 0.5) {$\kappa = 0$};

		 \draw[|-|, >=latex, thick, color = black] ($(levelB.east)+(-0.2+0.05,0.1+0.05)$) -- node [right = +0.0mm] {$\Delta$} ($(levelC.east)+(-0.2+0.05,-0.15)$);

		 \node at ($(levelC.north)+(-1.4, -1.25)$) {{\footnotesize Donor\phantom{p}}};
		 \node at ($(levelC.north)+(0.0, -1.25)$) {{\footnotesize Acceptor}};

	  \end{scope}

	  %%%%%%%%%%%%%%%%%%%%%%%%%%%%%%%%%
	  % second scheme with parabola
	  %%%%%%%%%%%%%%%%%%%%%%%%%%%%%%%%%

	  %\begin{scope}[yshift = -1.5cm,xshift = 4cm]
	  \begin{scope}[yshift = -1.1cm+1.0cm+1cm, xshift = 15.cm]

		 \draw[thick, line width = \mylevelw] (0.0,0) -- ++(1,0) node[midway, above=-2.5mm] (levelA) {};
		 \draw[thick, line width = \mylevelw, dash pattern=on 3pt off 3pt] (1.5, 0) -- ++(1,0) node[midway, above=-2.5mm] (levelB2) {};
		 \draw[<->, >=latex, thick, color = red] (levelA) edge[out=45,in=135] node [above = 0.25mm, color = black] {$J$} (levelB2);
		 \draw[thick, line width = \mylevelw, color = black] (1.5, 0.75) -- ++(1,0)  node[midway, below=-2.5mm] (levelC2) {};
		 %\draw[->, >=latex, thick] (levelA) edge[out=45,in=135] (levelB2);
		 \node at (-3.5, 0.5) {$\kappa > 0$};
		 \draw[<->, >=latex, thick, color = red] (levelB2) edge[out=75,in=-75] node[midway, below = -0.5mm] (uparrow) {} (levelC2);

		 % parabola
		 \begin{scope}[scale=0.25, line width = 0.5mm, xshift = 12.5cm, yshift = 0.5cm]
			\draw node[above left = 2.5mm] (myparabola) {} (-2,4) parabola bend (0,0) (2,4)  {};
			\node at (-1.8,0) {\large $\kappa$};
			\foreach \x in {1,2,3,4,5}
			{
			   \def\parfac{0.75}
			   \draw[thick, line width = 0.35mm] (-{sqrt(\parfac*\x)}, \parfac*\x) -- ({sqrt(\parfac*\x)},\parfac*\x) {};
			}
		 \end{scope}

		 \draw[>=latex, decorate, decoration={snake, amplitude = 1mm, segment length = 1.5mm, post length = 0mm}, draw=red,line width=0.4mm] (uparrow) -- ($(myparabola.east)+(-0.1,0.0)$);%(2,2);

		 %\node at ($(levelC2.north)+(-1.35, -1.25)$) {{\footnotesize donor\phantom{p}}};
		 %\node at ($(levelC2.north)+(0.0, -1.25)$) {{\footnotesize acceptor}};

	  \end{scope}

	  %\node at (-0.4, 1.2) {(a)};

   \end{scope}

\end{scope}

%%%%%%%%%%%%%%%%%%%%%%%%%%%%%%%%%
% pictures
%%%%%%%%%%%%%%%%%%%%%%%%%%%%%%%%%

\node[inner sep=0pt] (trap) at (-6.5,0.0) {\includegraphics[width=.25\textwidth]{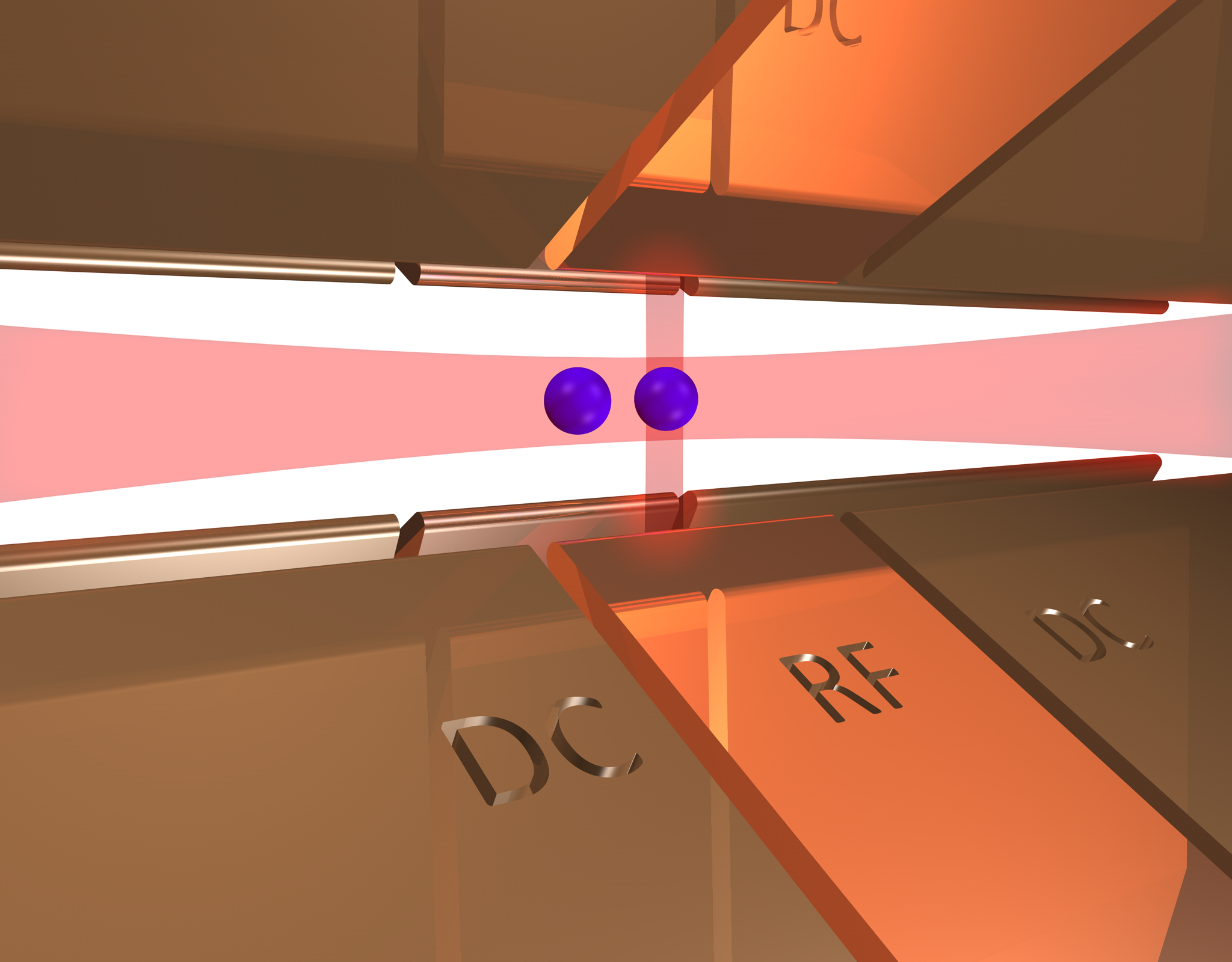}};

%\node at (-8.0, 0.3) {G};
%\node [scale = 0.75] at (-6.3, 0.6) {\footnotesize L};

%\node[inner sep=0pt] (scans) at (0.5-0.1+0.05-3.25, 4.0) {\includegraphics[width=0.325\textwidth]{fig1.eps}};
\node[inner sep=0pt] (scans) at (0.5-0.1+0.05-3.25, 4.0) {\includegraphics[width=0.325\textwidth]{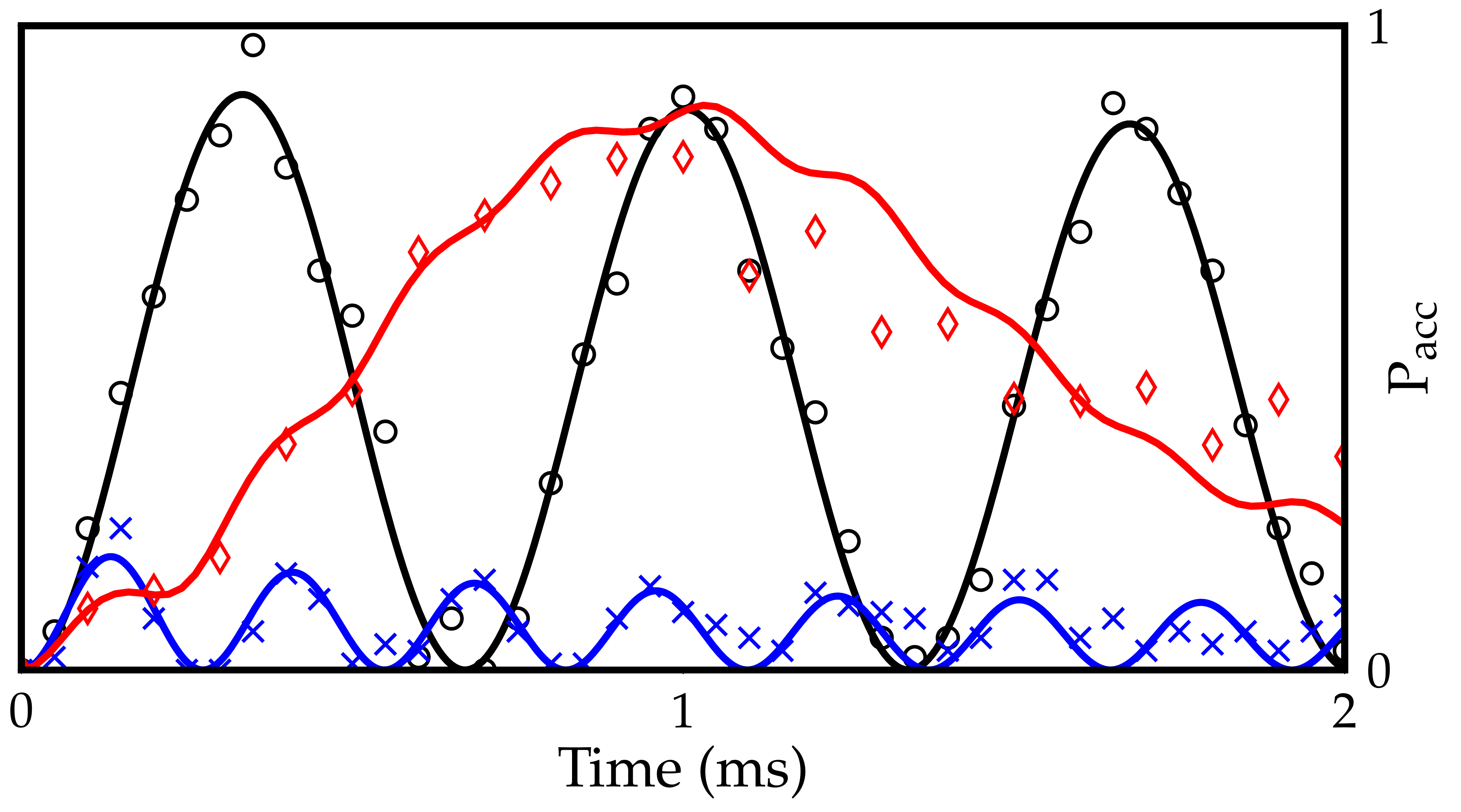}};

\node at (-8.6, 5.5) {(a)};
%\node at (3.3-0.15, 5.5) {(b)};
\node at (-8.6, 2) {(b)};
\node at (3.3-0.15, 2) {(c)};

\draw[->, >=latex, thick, color = blue] ($(levelC.east)+(1.5,-1.0-0.1)$) -- (-5.5+0.075-0.25,3.25);

%\draw[->, >=latex, thick, color = red] ($(levelC2.west)+(-1.5, -1.0)$) -- (-0.25,3.4);
%\draw[->, >=latex, thick, color = red] ($(levelC2.west)+(-1.5, -1.0-0.25)$) -- (-0.25-1.5,4.5);
\draw[->, >=latex, thick, color = red] ($(levelC2.west)+(-1.5-0.25, -1.0-0.25)$) -- ++(-3.0+0.25, 0.0);
%\draw[->, >=latex, thick, color = red] (1.0, 4.5+0.5) edge[out=180,in=75] (-1.0,4.5-0.5);

\draw[->, >=latex, thick, color = black] ($(levelC.east)+(1.25,0.15)$) -- (-5.25,4.5);

\begin{scope}[xshift = 0.5cm, yshift=-1.25cm, scale = 0.8]

   \begin{scope}

	  \draw[carrier, line width = \mylw] (0,0) -- ++(0,\carrierheight) {};

	  \foreach \sidespace in {\wspaceA, \wspaceB} {
		 \draw[rsb, line width = \mylw] (\sidespace,0) -- ++(0,\sbheight) {};
		 \draw[bsb, line width = \mylw] (-\sidespace,0) -- ++(0,\sbheight) {};
	  }

	  % MS gate
	  \foreach \x in {-0.8, 0.8} {
		 \draw[ms, line width = \mylw] (\x*\wspaceA,0) -- ++(0,\laserheight) {};	
	  }

	  \node at (-1.4*\wspaceB, 0.5*\sbheight) {Acceptor};

	  % baseline
	  \draw[->, >=latex, thick, line width = \mylw] (-\wspaceB * 1.15,0) -- ++(2*\wspaceB * 1.15,0) node [midway, below = 0.2mm] {Frequency} {};

   \end{scope}

   \begin{scope}[yshift = 1.75cm]

	  \draw[carrier, line width = \mylw] (0,0) -- ++(0,\carrierheight) {};

	  \foreach \sidespace in {\wspaceA, \wspaceB} {
		 \draw[rsb, line width = \mylw] (\sidespace,0) -- ++(0,\sbheight) {};
		 \draw[bsb, line width = \mylw] (-\sidespace,0) -- ++(0,\sbheight) {};
	  }

	  % MS gate
	  \foreach \x in {-0.8, 0.8} {
		 \draw[ms, line width = \mylw] (\x*\wspaceA,0) -- ++(0,\laserheight) {};	
	  }

	  % sigma_z gate
	  \draw[sigmaz, line width = \mylw] (-0.5*\wspaceB,0) -- ++(0,\laserheight) {};	
	  \draw[sigmaz, line width = \mylw] (+0.5*\wspaceB+0.2,0) -- ++(0,0.75*\laserheight) {};	

	  \node at (-1.4*\wspaceB, 0.5*\sbheight) {Donor};

	  % baseline
	  \draw[->, >=latex, thick, line width = \mylw] (-\wspaceB * 1.15,0) -- ++(2*\wspaceB * 1.15,0) {};

	  % sigma_z beams
	  \draw[<->, >=latex, thick, line width = \mylw] (-0.5 * \wspaceB,1.35) -- ++(\wspaceB + 0.2,0) {};
	  %\draw[carrier, line width = \mylw,  dash pattern=on 3pt off 2.5pt] (0.5 * \wspaceB, 0) -- ++(0,\carrierheight) {};

	  % text
	  \node at (0, 1.6) {$\omega_r + \nu_{\rm{eff}}$};
	  \node at (\wspaceA + 0.2, -0.25) {$\omega_\textrm{ax}$};
	  \node at (\wspaceB + 0.075, -0.25) {$\omega_r$};

	  \node at (-\wspaceA - 0.2, -0.25) {$-\omega_\textrm{ax}$};
	  \node at (-\wspaceB - 0.075, -0.25) {$-\omega_r$};

   \end{scope}

\end{scope}

\end{tikzpicture}

%% file: fig2box.tex
\begin{tikzpicture}[scale = 1.0]

\node[inner sep=0pt] (scans) at (5.5+0.75, -0.2) {\includegraphics[width=0.37\textwidth]{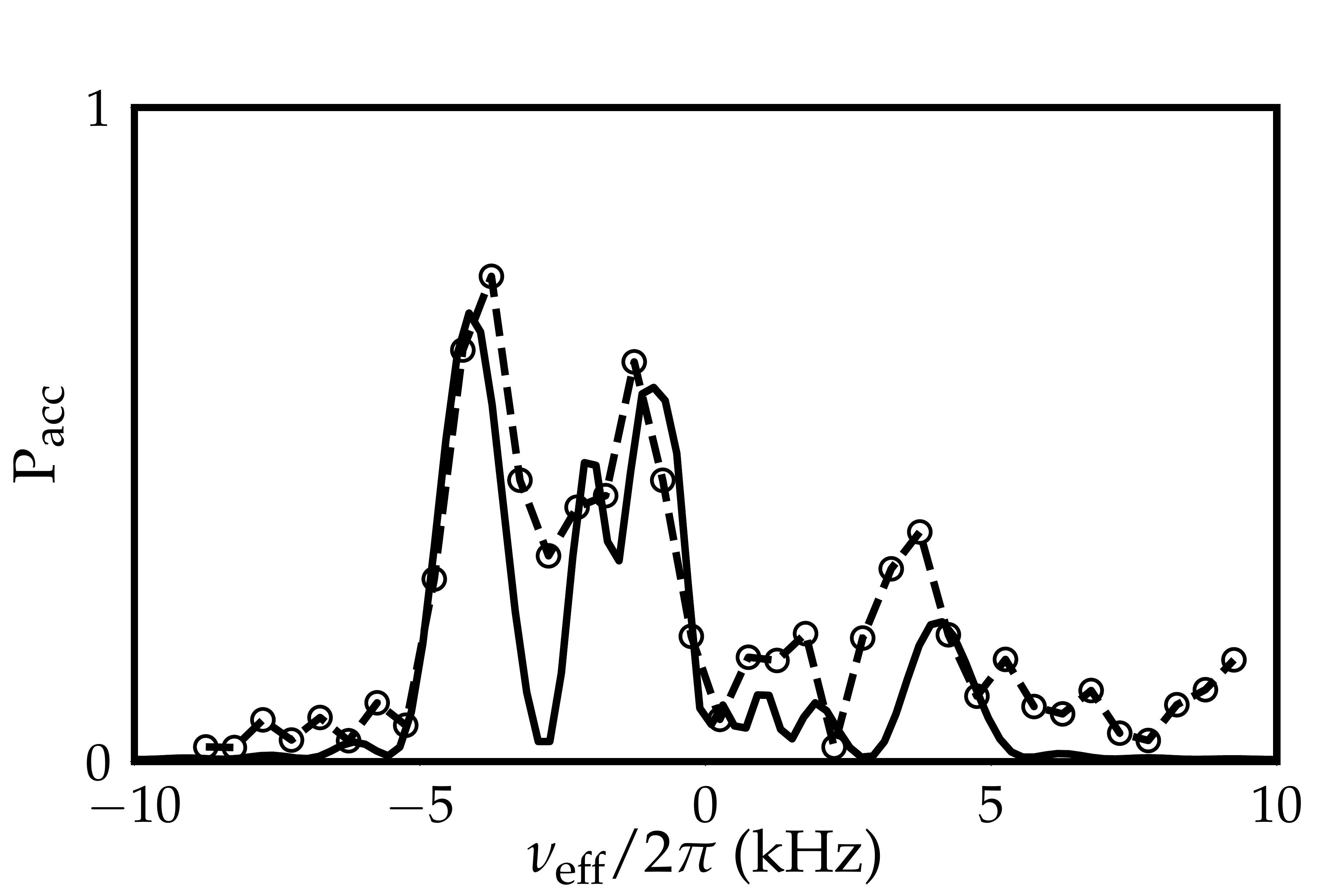}};
\node[inner sep=0pt] (scans) at (5.5+0.75, 4.5-0.5) {\includegraphics[width=0.37\textwidth]{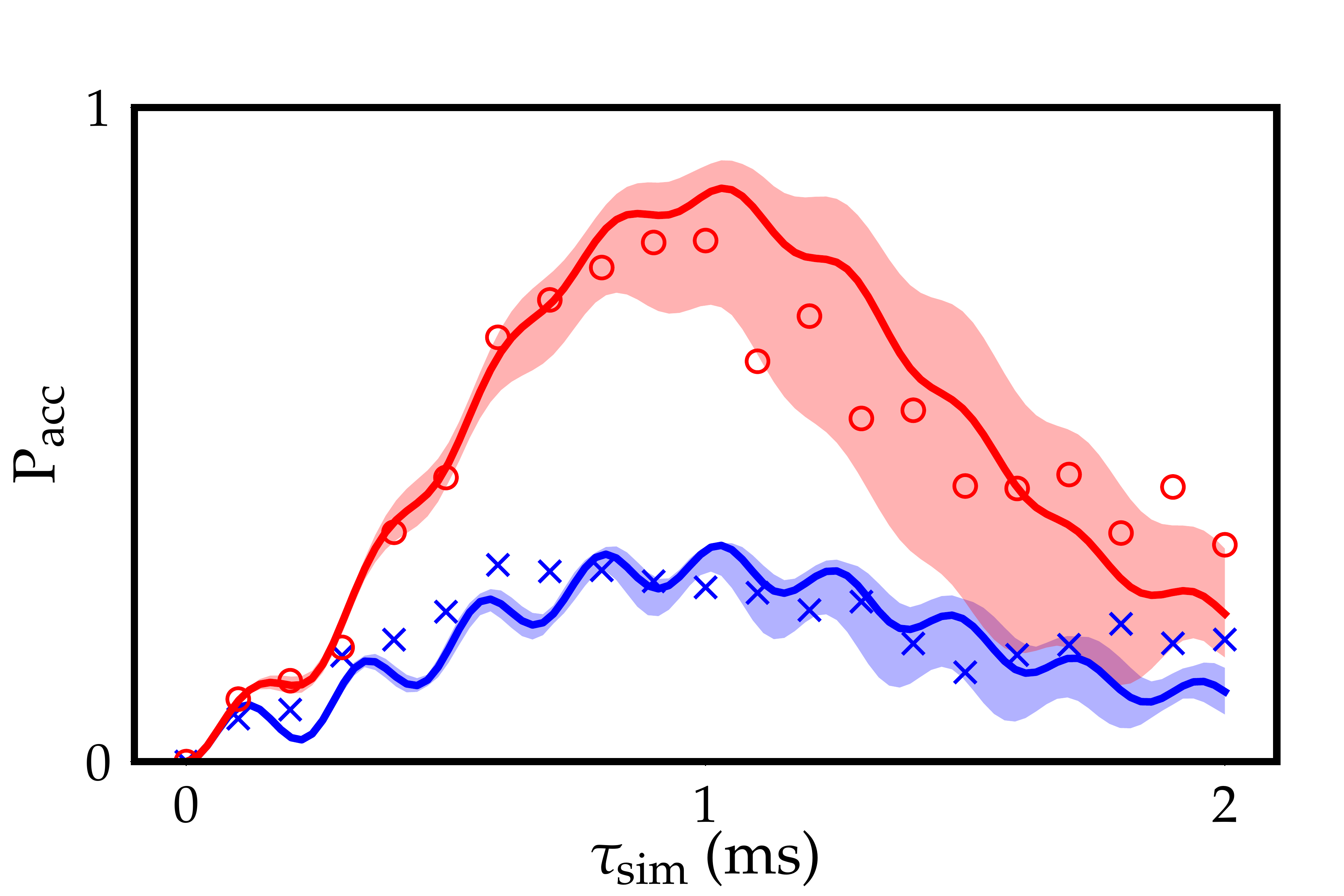}};

\node[inner sep=0pt] (trap) at (0.0, -0.2) {\includegraphics[width=0.37\textwidth]{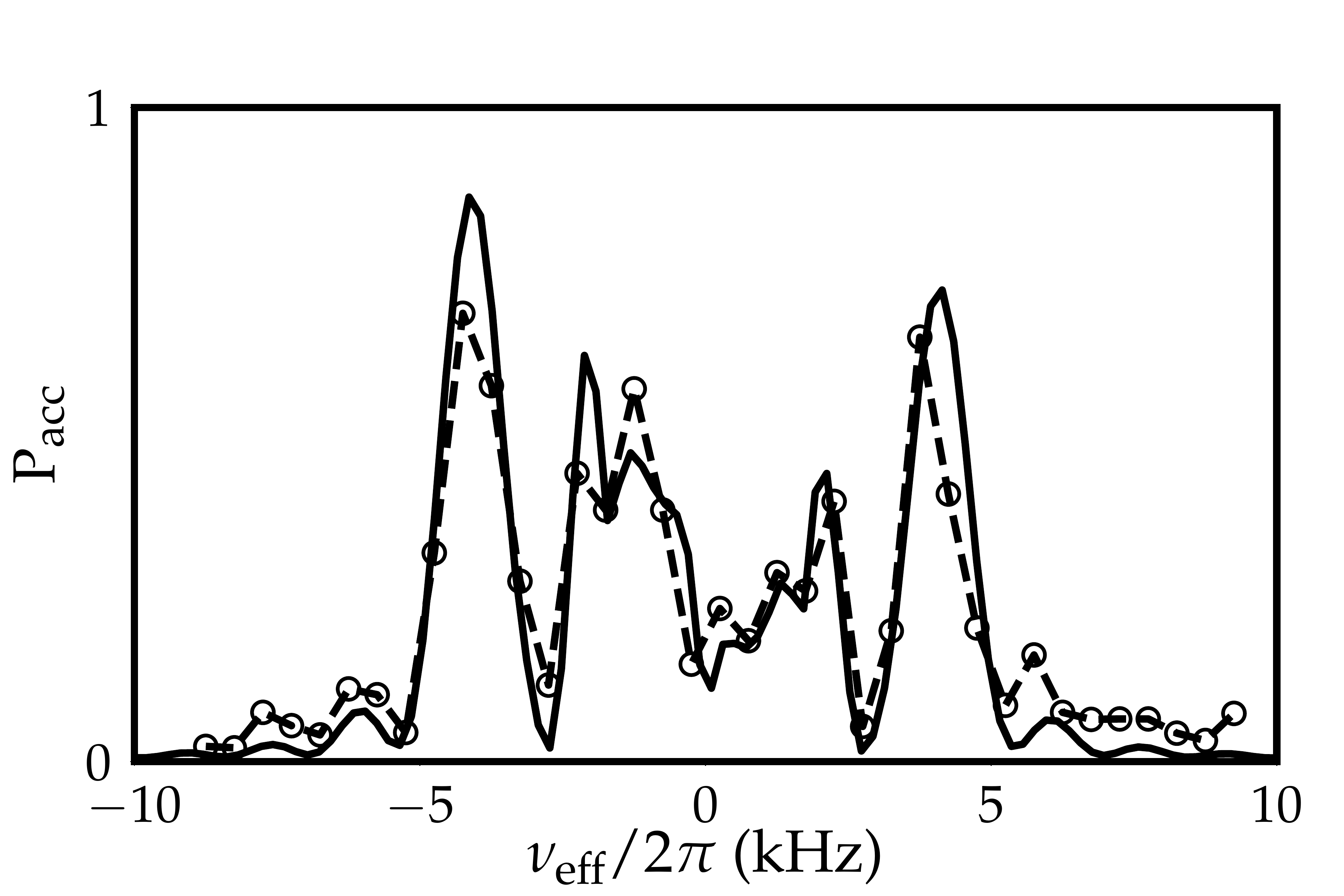}};
\node[inner sep=0pt] (trap) at (0.0, 4.5-0.5) {\includegraphics[width=0.37\textwidth]{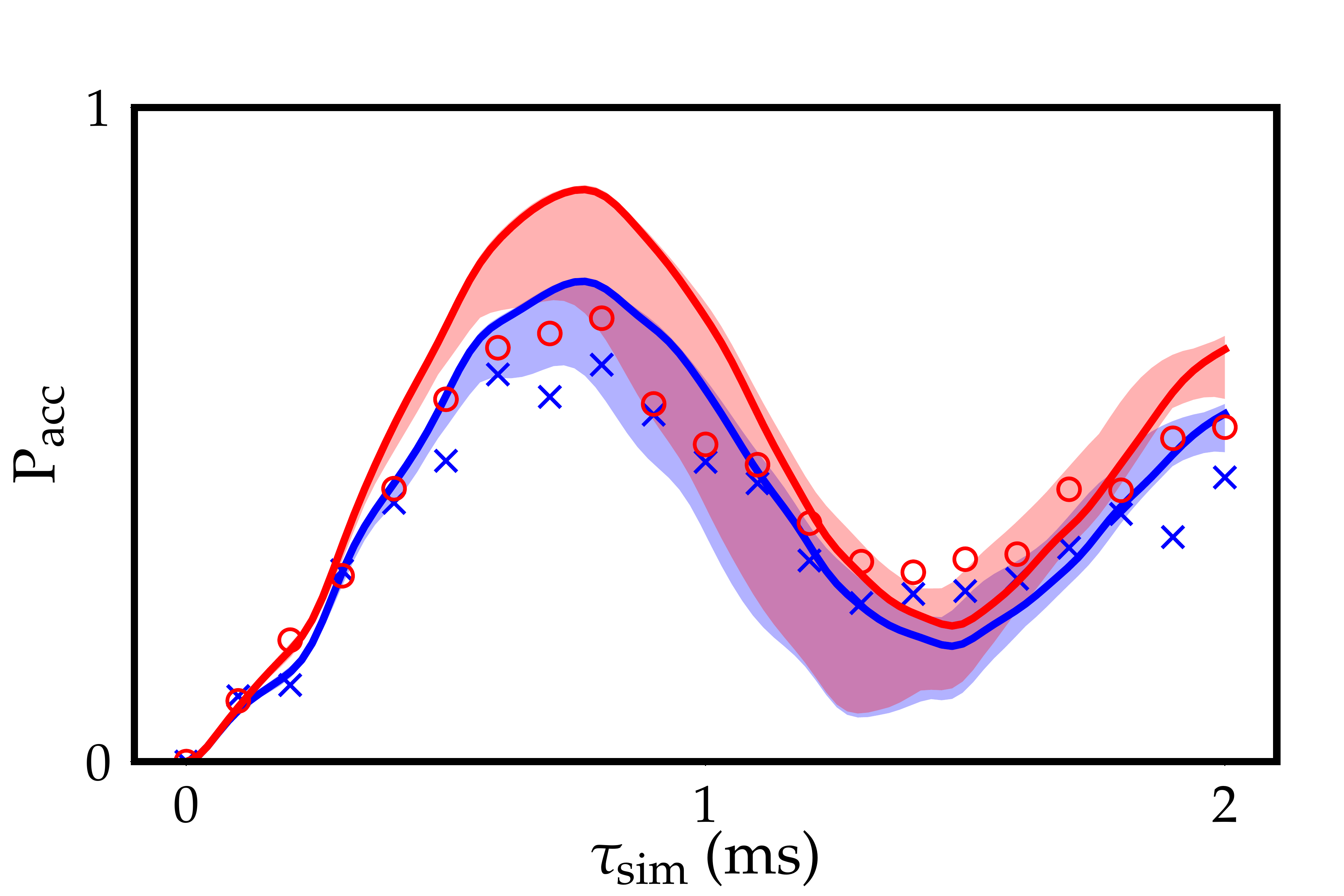}};

\fill [white] (-3.0+0.36,1.5) rectangle (3.0,1.7);
\fill [white] (3.0-0.05,1.7) rectangle (3.1,-2+0.25);

\def\offset{6.0+0.25}

\fill [white] (-3.0+0.36 + \offset,1.5) rectangle (3.0 + \offset,1.7);
\fill [white] (3.0-0.05 + \offset,1.7) rectangle (3.1 + \offset,-2+0.25);

\draw[<-, >=latex, thick, color = blue, line width = 0.5mm] (1.1,3.8-0.0-0.7-0.2) -- (1.1+0.15,0.95);
\draw[<-, >=latex, thick, color = red, line width = 0.5mm] (-1.6,4.1-0.5-0.4) -- (-1.0,1.3-0.1);

\draw[<-, >=latex, thick, color = blue, line width = 0.5mm] (6.5+0.75,3.875-0.75) -- (6.45+0.75+0.2,-0.15);
\draw[<-, >=latex, thick, color = red, line width = 0.5mm] (5+0.75+0.1,5.5-0.55-0.15) -- (4.585+0.75,0.85);

%\node [
%draw = black,
%left color = red,
%right color=blue,
%    single arrow,
%    minimum height=1.5cm, scale = 1
%] at (3., 3.25-1-1-0.25) {};

%\node [scale=0.6] at (3., 3.75-1) {T lower};

%\node at (-2-0.15, 6.05-0.1-0.7) {(a)};
%\node at (7.65+0.75+0.5, 6.05-0.1-0.7) {(b)};

\node at (-2-0.15 - 1.0, 1.5-0.2 + 4.0 + 0.5) {(a)};
\node at (7.65+0.75+0.5 + 0.75 - 6.0 + 1.5 + 4.5, 1.5-0.2 + 4.0 + 0.5) {(b)};

\node at (0, 6) {High T: $\bar{n} = 5$};
\node at (5.5+0.75, 6) {Low T: $\bar{n} = 0.5$};

%\draw[draw, color = black, line width = 1pt, rounded corners, rounded corners = 3pt, scale = 1.0] (-3, -2.2) rectangle (8.5, 6.5) {};

\end{tikzpicture}

%% file: fig3box.tex
\begin{tikzpicture}

   \begin{scope}
	  
%	  \node[inner sep=0pt] (trap) at (5.0+0.25-0.5, 0.0) {\includegraphics[width=0.5\textwidth]{fig3_b.eps}};
%	  \node[inner sep=0pt] (trap) at (-2.0+0.2-0.25-2.0, 0.0) {\includegraphics[width=0.5\textwidth]{fig3_a.eps}};

	  \node[inner sep=0pt] (trap) at (5.0+0.25-0.5, 0.0) {\includegraphics[width=0.5\textwidth]{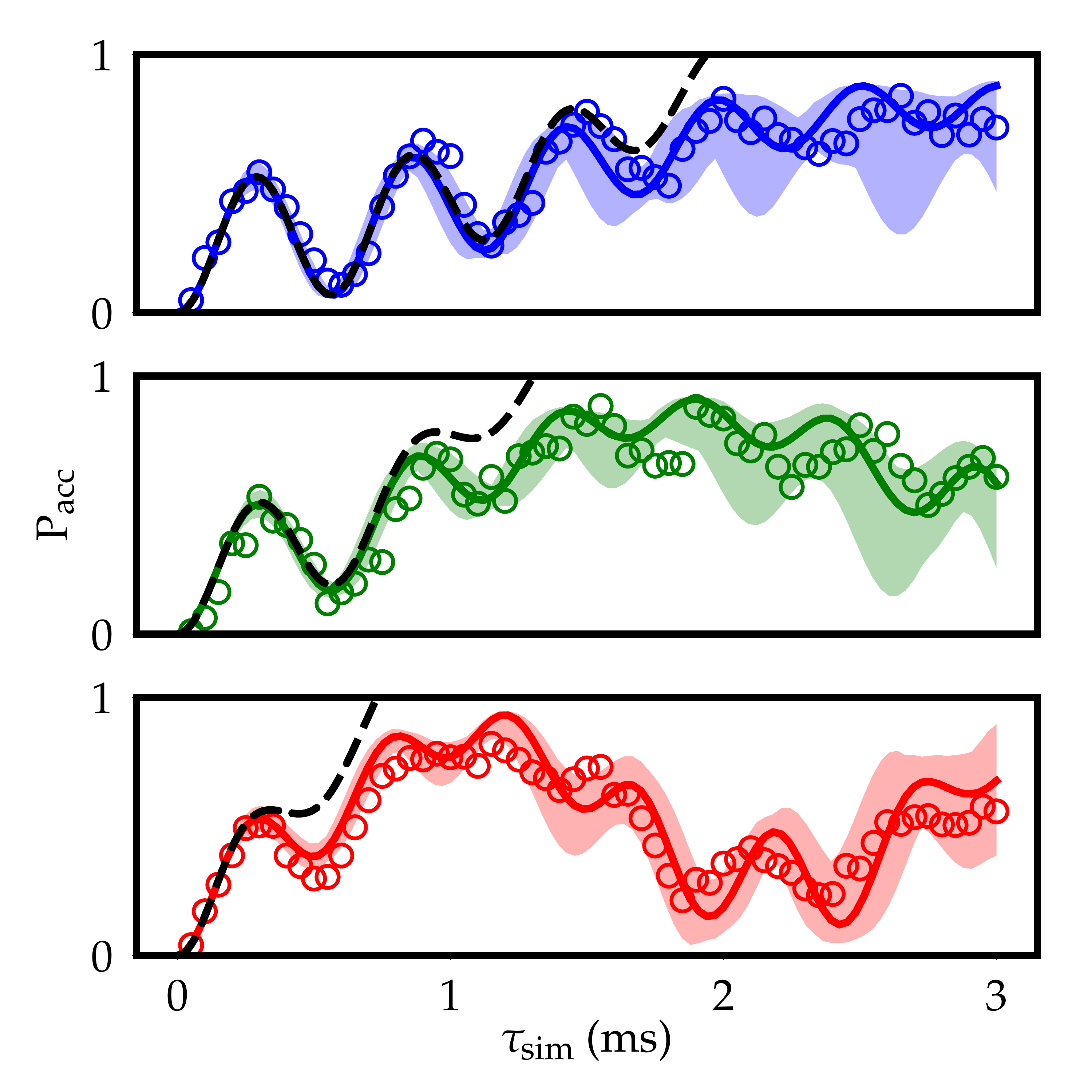}};
	  \node[inner sep=0pt] (trap) at (-2.0+0.2-0.25-2.0, 0.0) {\includegraphics[width=0.5\textwidth]{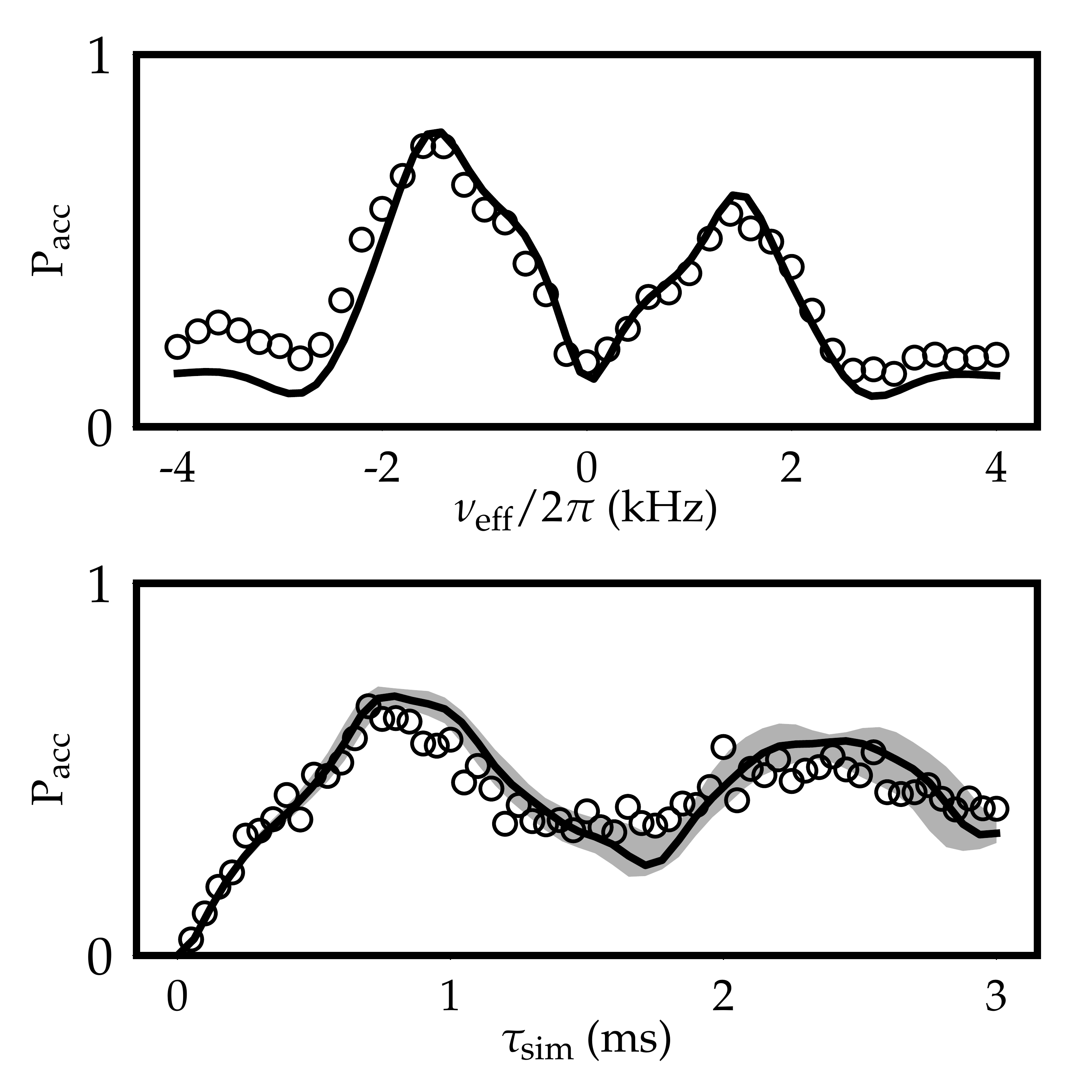}};

    \node [draw, thick, shape=rectangle, minimum width=250, minimum height=260, anchor=center] at (5.0+0.25-0.5, 0.0) {};	  

	  \node at (-4.25-0.25-0.25-2.25, 3+0.25+0.1+0.3) {(a)};
	  \node at (-4.25-0.25-0.25-2.25, -0.75+0.25-0.1+0.0) {(c)};
	  \node at (8.0+0.25 - 5.4 - 0.25-0.75 - 1.2,  3.0+0.25+0.1+0.3+0.3) {(b)};

	  \node [scale = 0.85] at (8.0 - 1.0 - 0.15 + 0.2, 2.0-0.25+0.1+0.15+0.25) {$\kappa = 2\pi \times 0.229(6)$ kHz};
	  \node [scale = 0.85] at (8.0 - 1.0 - 0.1 + 0.2, 2.0-0.25+0.1 - 2.1 - 0.225) {$\kappa = 2\pi \times 0.37(1)$ kHz};
	  \node [scale = 0.85] at (8.0 - 1.0 - 0.1 + 0.1, 2.0-0.25+0.1 - 3.1-0.1-0.05-0.1) {$\kappa = 2\pi \times 0.64(2)$ kHz};

   \end{scope}

\end{tikzpicture}

%% file: supplfig.tex
\begin{tikzpicture}[scale = 1.0]
\node[inner sep=0pt] (trap) at (-1.0,  0.0) {\includegraphics[width=0.45\textwidth]{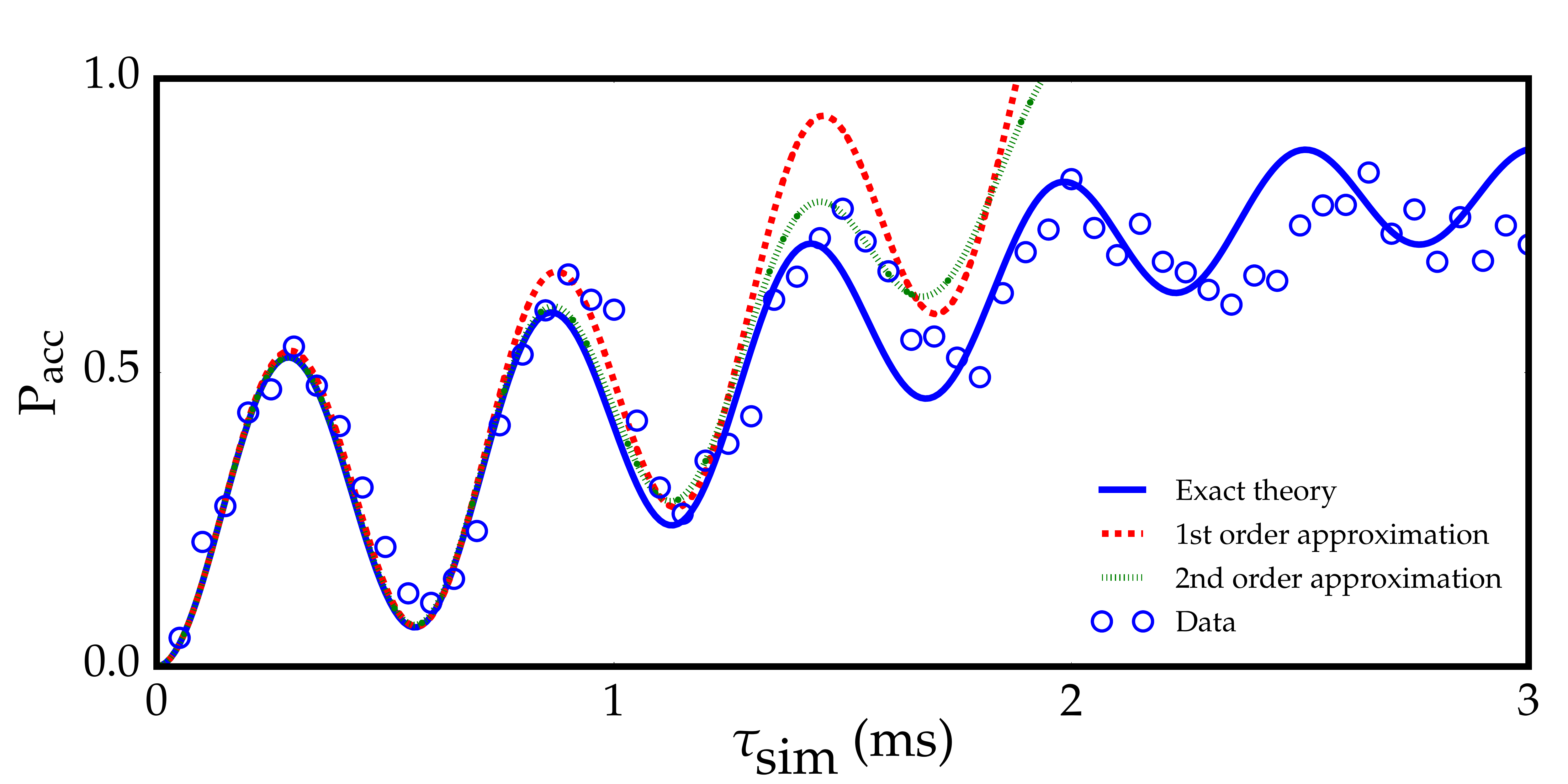}};
\node[inner sep=0pt] (trap) at (7.5,  0.0) {\includegraphics[width=0.45\textwidth]{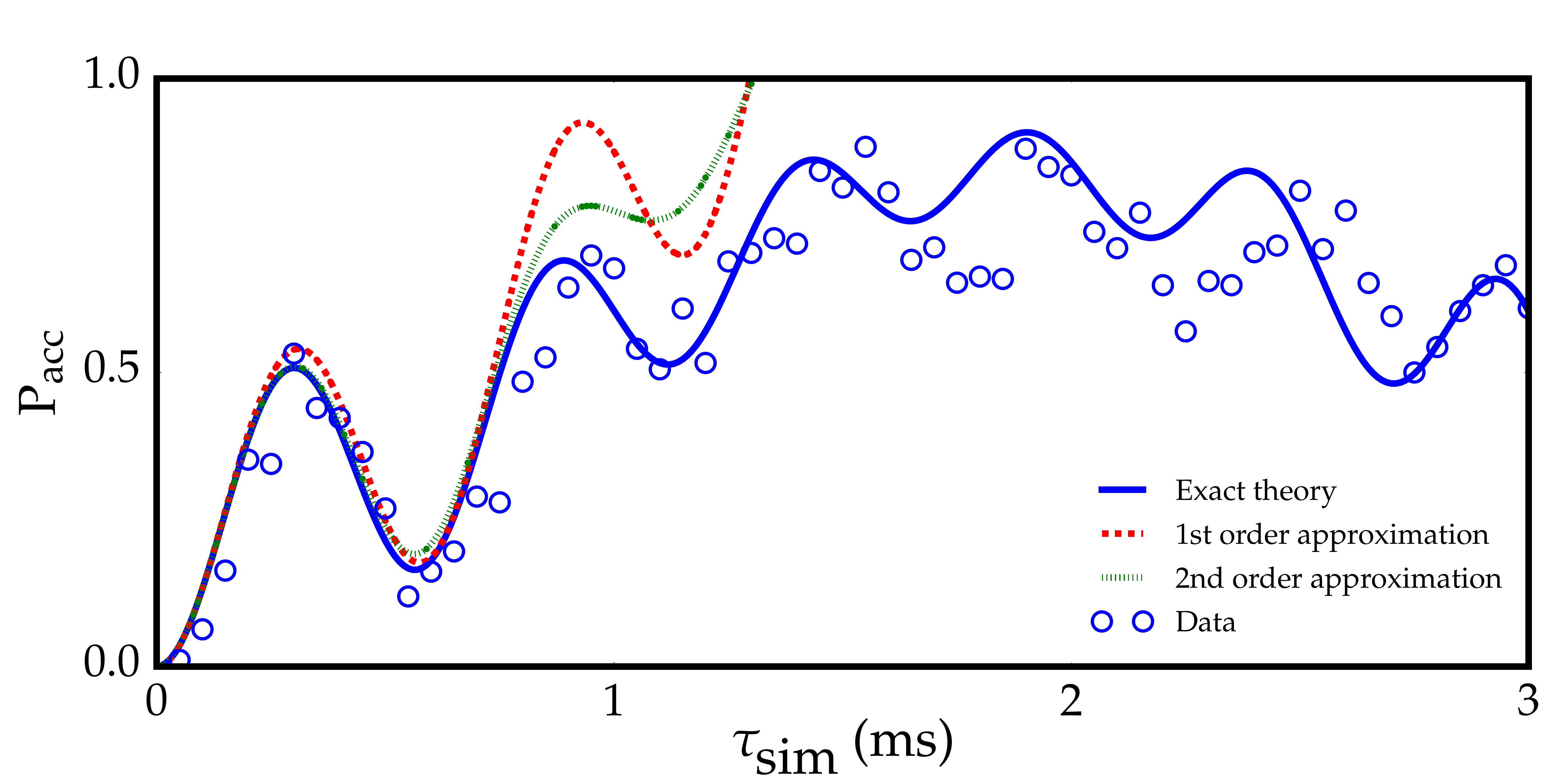}};
\node[inner sep=0pt] (trap) at (-1.0, -4.0) {\includegraphics[width=0.45\textwidth]{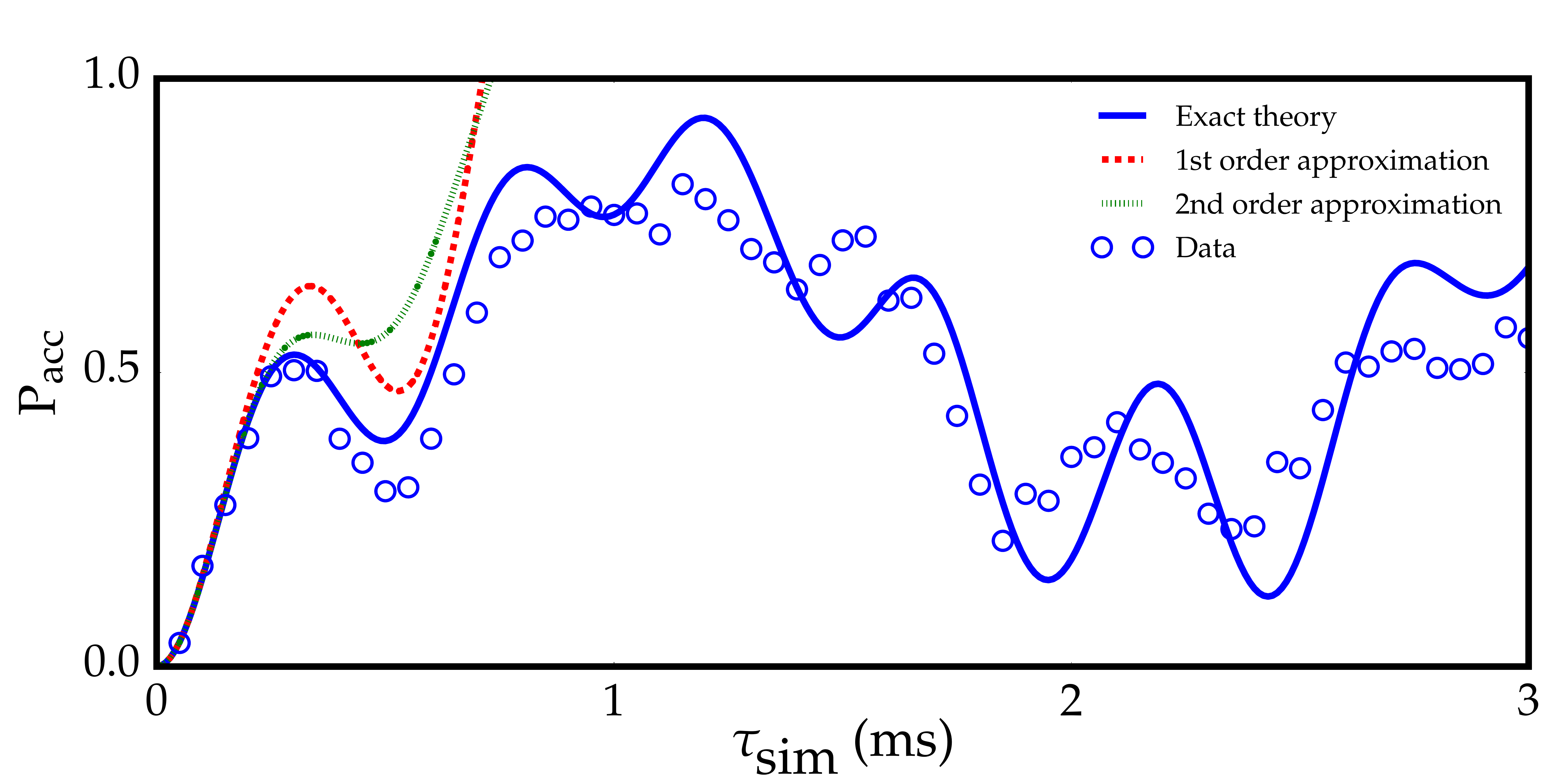}};

%\draw[draw, color = black, line width = 1pt, rounded corners, rounded corners = 3pt, scale = 1.0] (-7.4,-2.0) rectangle (1.4,2.25) {};

%\draw [line width = 1pt] (-3-0.2, -2.0) -- (-3-0.2, 2.25) {};

%\node at (-6.7, 1.55) {(a)};
%\node at (-6.7, -0.35) {(b)};
%\node at (0.9,  1.15) {(c)};

\end{tikzpicture}